\documentclass{emulateapj}
\pdfoutput=1
\usepackage{epstopdf}
\usepackage{natbib}
\usepackage{amsmath}

\DeclareGraphicsExtensions{.jpg,.pdf,.png,.eps,.ps}

\newcommand{\kms}{\,km\,s$^{-1}$}

\newcommand{\sqdeg}{\protect\hbox{deg$^2$} }

\newcommand{\msun}{$M_{\sun}$}

\def\lsim{\hbox{\rlap{\raise 0.425ex\hbox{$<$}}\lower 0.65ex\hbox{$\sim$}}}
\def\gsim{\hbox{\rlap{\raise 0.425ex\hbox{$>$}}\lower 0.65ex\hbox{$\sim$}}}
\def\arcmin{\hbox{$^\prime$}}
\def\arcsec{\hbox{$^{\prime\prime}$}}

\newcommand{\cluster}{SPT-CL~J2040-4451}

\newcommand{\zp}{$z = 1.37 \pm 0.07$}

\newcommand{\zsnoe}{$z = 1.478$}

\begin{document}

\title{\cluster: An SZ-Selected Galaxy Cluster at \lowercase{$z = 1.478$} With Significant Ongoing Star Formation} 
%Discovered By the South Pole Telescope}

\def\Harvard{1}
\def\CfA{2}
\def\Miss{3}
\def\UChicago{4}
\def\MIT{5}
\def\KICPChicago{6}
\def\EFIChicago{7}
\def\PhysicsUChicago{8}
\def\ANL{9}
\def\Munich{10}
\def\ExcellenceCluster{11}
\def\AAUChicago{12}
\def\NIST{13}
\def\PUC{14}
\def\McGill{15}
\def\UIastro{16}
\def\UIphysics{17}
\def\Berkeley{18}
\def\UFlorida{19}
\def\Colorado{20}
\def\NASA{21}
\def\Davis{22}
\def\LBNL{23}
\def\Caltech{24}
\def\Arizona{25}
\def\Michigan{26}
\def\MPE{27}
\def\CaseWestern{28}
\def\Minnesota{29}
\def\STScI{30}
\def\SAIC{31}
\def\LLNL{32}
\def\Dunlap{33}
\def\UofT{34}
\def\BCCP{35}
\def\hubble{*}

\altaffiltext{\Harvard}{Department of Physics, Harvard University, 17 Oxford Street, Cambridge, MA 02138}
\altaffiltext{\CfA}{Harvard-Smithsonian Center for Astrophysics, 60 Garden Street, Cambridge, MA 02138}
\altaffiltext{\Miss}{Department of Physics and Astronomy, University of Missouri, 5110 Rockhill Road, Kansas City, MO 64110}
\altaffiltext{\UChicago}{University of Chicago, 5640 South Ellis Avenue, Chicago, IL 60637}
\altaffiltext{\MIT}{Kavli Institute for Astrophysics and Space Research, Massachusetts Institute of Technology, 77 Massachusetts Avenue, Cambridge, MA 02139}
\altaffiltext{\KICPChicago}{Kavli Institute for Cosmological Physics, University of Chicago, 5640 South Ellis Avenue, Chicago, IL 60637}
\altaffiltext{\EFIChicago}{Enrico Fermi Institute, University of Chicago, 5640 South Ellis Avenue, Chicago, IL 60637}
\altaffiltext{\PhysicsUChicago}{Department of Physics, University of Chicago, 5640 South Ellis Avenue, Chicago, IL 60637}
\altaffiltext{\ANL}{Argonne National Laboratory, 9700 S. Cass Avenue, Argonne, IL, USA 60439}
\altaffiltext{\Munich}{Department of Physics, Ludwig-Maximilians-Universit\"{a}t, Scheinerstr.\ 1, 81679 M\"{u}nchen, Germany}
\altaffiltext{\ExcellenceCluster}{Excellence Cluster Universe, Boltzmannstr.\ 2, 85748 Garching, Germany}
\altaffiltext{\AAUChicago}{Department of Astronomy and Astrophysics, University of Chicago, 5640 South Ellis Avenue, Chicago, IL 60637}
\altaffiltext{\NIST}{NIST Quantum Devices Group, 325 Broadway Mailcode 817.03, Boulder, CO, USA 80305}
\altaffiltext{\PUC}{Departamento de Astronomia y Astrosifica, Pontificia Universidad Catolica, Chile}
\altaffiltext{\McGill}{Department of Physics, McGill University, 3600 Rue University, Montreal, Quebec H3A 2T8, Canada}
\altaffiltext{\UIastro}{Astronomy Department, University of Illinois at Urbana-Champaign, 1002 W. Green St., Urbana, IL 61801}
\altaffiltext{\UIphysics}{Department of Physics, University of Illinois at Urbana-Champaign, 1110 W. Green St., Urbana, IL 61801}
\altaffiltext{\Berkeley}{Department of Physics, University of California, Berkeley, CA 94720}
\altaffiltext{\UFlorida}{Department of Astronomy, University of Florida, Gainesville, FL 32611}
\altaffiltext{\Colorado}{Department of Astrophysical and Planetary Sciences and Department of Physics, University of Colorado, Boulder, CO 80309}
\altaffiltext{\NASA}{Department of Space Science, VP62, NASA Marshall Space Flight Center, Huntsville, AL 35812}
\altaffiltext{\Davis}{Department of Physics, University of California, One Shields Avenue, Davis, CA 95616}
\altaffiltext{\LBNL}{Physics Division, Lawrence Berkeley National Laboratory, Berkeley, CA 94720}
\altaffiltext{\Caltech}{California Institute of Technology, 1200 E. California Blvd., Pasadena, CA 91125}
\altaffiltext{\Arizona}{Steward Observatory, University of Arizona, 933 North Cherry Avenue, Tucson, AZ 85721}
\altaffiltext{\Michigan}{Department of Physics, University of Michigan, 450 Church Street, Ann  Arbor, MI, 48109}
\altaffiltext{\MPE}{Max-Planck-Institut f\"{u}r extraterrestrische Physik, Giessenbachstr.\ 85748 Garching, Germany}
\altaffiltext{\CaseWestern}{Physics Department, Center for Education and Research in Cosmology and Astrophysics, Case Western Reserve University, Cleveland, OH 44106}
\altaffiltext{\Minnesota}{Physics Department, University of Minnesota, 116 Church Street S.E., Minneapolis, MN 55455}
\altaffiltext{\STScI}{Space Telescope Science Institute, 3700 San Martin Dr., Baltimore, MD 21218}

\altaffiltext{\SAIC}{Liberal Arts Department, School of the Art Institute of Chicago, 112 S Michigan Ave, Chicago, IL 60603}
\altaffiltext{\LLNL}{Institute of Geophysics and Planetary Physics, Lawrence Livermore National Laboratory, Livermore, CA 94551}
\altaffiltext{\Dunlap}{Dunlap Institute for Astronomy \& Astrophysics, University of Toronto, 50 St George St, Toronto, ON, M5S 3H4, Canada}
\altaffiltext{\UofT}{Department of Astronomy \& Astrophysics, University of Toronto, 50 St George St, Toronto, ON, M5S 3H4, Canada}
\altaffiltext{\BCCP}{Berkeley Center for Cosmological Physics, Department of Physics, University of California, and Lawrence Berkeley National Labs, Berkeley, CA 94720}
\altaffiltext{\hubble}{Hubble Fellow}

%\author{You and Me and Everyone We know}
\author{M.~B.~Bayliss\altaffilmark{\Harvard,\CfA}, 
M.~L.~N.~Ashby\altaffilmark{\CfA},
J.~Ruel\altaffilmark{\Harvard},
M.~Brodwin\altaffilmark{\Miss},
K.~A.~Aird\altaffilmark{\UChicago},
M.~W.~Bautz\altaffilmark{\MIT},
B.~A.~Benson\altaffilmark{\KICPChicago,\EFIChicago},
L.~E.~Bleem\altaffilmark{\KICPChicago,\PhysicsUChicago,\ANL},
S.~Bocquet\altaffilmark{\Munich,\ExcellenceCluster},
J.~E.~Carlstrom\altaffilmark{\KICPChicago,\EFIChicago,\PhysicsUChicago,\ANL,\AAUChicago}, 
C.~L.~Chang\altaffilmark{\KICPChicago,\EFIChicago,\ANL}, 
H.~M. Cho\altaffilmark{\NIST}, 
A.~Clocchiatti\altaffilmark{\PUC},
T.~M.~Crawford\altaffilmark{\KICPChicago,\AAUChicago},
A.~T.~Crites\altaffilmark{\KICPChicago,\AAUChicago},
S.~Desai\altaffilmark{\Munich,\ExcellenceCluster},
M.~A.~Dobbs\altaffilmark{\McGill},
J.~P.~Dudley\altaffilmark{\McGill},
R.~J.~Foley\altaffilmark{\CfA,\UIastro,\UIphysics}, 
W.~R.~Forman\altaffilmark{\CfA},
E.~M.~George\altaffilmark{\Berkeley},
D.~Gettings\altaffilmark{\UFlorida},
M.~D.~Gladders\altaffilmark{\KICPChicago,\AAUChicago},
A.~H.~Gonzalez\altaffilmark{\UFlorida},
T.~de~Haan\altaffilmark{\McGill},
N.~W.~Halverson\altaffilmark{\Colorado},
F.~W.~High\altaffilmark{\KICPChicago,\AAUChicago}, 
G.~P.~Holder\altaffilmark{\McGill},
W.~L.~Holzapfel\altaffilmark{\Berkeley},
S.~Hoover\altaffilmark{\KICPChicago,\EFIChicago},
J.~D.~Hrubes\altaffilmark{\UChicago},
C.~Jones\altaffilmark{\CfA},
M.~Joy\altaffilmark{\NASA},
R.~Keisler\altaffilmark{\KICPChicago,\PhysicsUChicago},
L.~Knox\altaffilmark{\Davis},
A.~T.~Lee\altaffilmark{\Berkeley,\LBNL},
E.~M.~Leitch\altaffilmark{\KICPChicago,\AAUChicago},
J.~Liu\altaffilmark{\Munich,\ExcellenceCluster},
M.~Lueker\altaffilmark{\Berkeley,\Caltech},
D.~Luong-Van\altaffilmark{\UChicago},
A.~Mantz\altaffilmark{\KICPChicago},
D.~P.~Marrone\altaffilmark{\Arizona},
K.~Mawatari\altaffilmark{\CfA},
M.~McDonald\altaffilmark{\MIT,\hubble},
J.~J.~McMahon\altaffilmark{\Michigan},
J.~Mehl\altaffilmark{\KICPChicago,\AAUChicago,\ANL},
S.~S.~Meyer\altaffilmark{\KICPChicago,\EFIChicago,\PhysicsUChicago,\AAUChicago},
E.~D.~Miller\altaffilmark{\MIT},
L.~Mocanu\altaffilmark{\KICPChicago,\AAUChicago},
J.~J.~Mohr\altaffilmark{\Munich,\ExcellenceCluster,\MPE},
T.~E.~Montroy\altaffilmark{\CaseWestern},
S.~S.~Murray\altaffilmark{\CfA},
S.~Padin\altaffilmark{\KICPChicago,\AAUChicago,\Caltech},
T.~Plagge\altaffilmark{\KICPChicago,\AAUChicago},
C.~Pryke\altaffilmark{\Minnesota}, 
C.~L.~Reichardt\altaffilmark{\Berkeley},
A.~Rest\altaffilmark{\STScI},
J.~E.~Ruhl\altaffilmark{\CaseWestern}, 
B.~R.~Saliwanchik\altaffilmark{\CaseWestern}, 
A.~Saro\altaffilmark{\Munich},
J.~T.~Sayre\altaffilmark{\CaseWestern}, 
K.~K.~Schaffer\altaffilmark{\KICPChicago,\EFIChicago,\SAIC}, 
E.~Shirokoff\altaffilmark{\Berkeley,\Caltech}, 
J.~Song\altaffilmark{\Michigan},
B.~Stalder\altaffilmark{\CfA},
R.~\v{S}uhada\altaffilmark{\Munich},
H.~G.~Spieler\altaffilmark{\LBNL},
S.~A.~Stanford\altaffilmark{\Davis,\LLNL},
Z.~Staniszewski\altaffilmark{\CaseWestern},
A.~A.~Stark\altaffilmark{\CfA}, 
K.~Story\altaffilmark{\KICPChicago,\PhysicsUChicago},
C.~W.~Stubbs\altaffilmark{\Harvard,\CfA}, 
A.~van~Engelen\altaffilmark{\McGill},
K.~Vanderlinde\altaffilmark{\Dunlap,\UofT},
J.~D.~Vieira\altaffilmark{\KICPChicago,\PhysicsUChicago,\UIastro,\UIphysics,\Caltech},
A. Vikhlinin\altaffilmark{\CfA},
R.~Williamson\altaffilmark{\KICPChicago,\AAUChicago}, 
O.~Zahn\altaffilmark{\Berkeley,\BCCP}, and
A.~Zenteno\altaffilmark{\Munich,\ExcellenceCluster}
}

\email{mbayliss@cfa.harvard.edu}

%\slugcomment{Submitted to \apj}

\begin{abstract}

\cluster~-- spectroscopically confirmed at \lowercase{\zsnoe} -- is the highest redshift 
galaxy cluster yet discovered via the Sunyaev-Zel'dovich effect. \cluster~was a 
candidate galaxy cluster identified in the first 720 deg$^{2}$ of the South Pole Telescope 
Sunyaev-Zel'dovich (SPT-SZ) survey, and has been confirmed in follow-up imaging and 
spectroscopy. From multi-object spectroscopy with Magellan-I/Baade$+$IMACS we measure 
spectroscopic redshifts for 15 cluster member galaxies, all of which have strong [O II]$\lambda\lambda$3727 
emission. \cluster~has an SZ-measured mass of M$_{500,SZ}=$ 3.2 $\pm$ 0.8 
$\times$ 10$^{14}$ M$_{\odot}$ h$_{70}^{-1}$, corresponding to M$_{200,SZ} =$ 5.8 
$\pm$~1.4~$\times$~10$^{14}$ M$_{\odot}$ h$_{70}^{-1}$. The velocity 
dispersion measured entirely from blue star forming members is $\sigma_{v} =$ 
1500 $\pm$ 520 km s$^{-1}$. The prevalence of star forming cluster members 
(galaxies with $>$ 1.5 M$_{\odot}$ yr$^{-1}$) implies that this massive, high-redshift 
cluster is experiencing a phase of active star formation, and supports recent results 
showing a marked increase in star formation occurring in galaxy clusters at 
$z \gtrsim 1.4$. We also compute the probability of finding a cluster as rare as this 
in the SPT-SZ survey to be $> 99\%$, indicating that its discovery is not in tension 
with the concordance $\Lambda$CDM cosmological model.  
\end{abstract}


%%%%%%%%%%%%%%%%%%%%
%%  Introduction  %%
%%%%%%%%%%%%%%%%%%%%

\section{Introduction}

As the most massive collapsed structures in the universe, galaxy 
clusters are both a sensitive probe of cosmology and an extreme 
environment for studying galaxy evolution. Specifically, galaxy clusters 
are the most over-dense environments in the universe and provide a 
laboratory for constraining the astrophysics of how galaxies form stars 
and evolve 
\citep[e.g.,][]{oemler74,dressler80,dressler83,balogh97,blanton09}. 
Massive galaxy clusters evolve from the most extreme peaks of the initial 
cosmic matter distribution, and until recently there was a consensus in the literature 
that the galaxies in clusters formed during a short-lived burst of star formation at 
early times ($z \gtrsim 3$) before quickly settling into a stable mode of passive 
evolution \citep{stanford98,holden05,stanford05,mei06}. However, recent studies 
of clusters at $z > 1$ have begun to reveal evidence for an era of active star 
formation and evolution of the cluster luminosity function at $z \gtrsim 1.4$ 
\citep{hilton09,mancone10,tran10,fassbender11,mancone12,snyder12,zeimann12,brodwin13}, 
suggesting that clusters in this epoch of the universe are undergoing a phase of 
significant galaxy assembly.

The high redshift frontier for both cosmological and astrophysical studies of galaxy 
clusters is now extended well beyond $z \gtrsim 1$, where large, well-defined 
samples of galaxy clusters have only recently begun to emerge. Several groups 
have had success identifying high-redshift galaxy clusters using deep observations 
at X-ray \citep[e.g.,][]{rosati04, mullis05, stanford06, rosati09} and optical$+$near 
infrared (NIR) wavelengths \citep[e.g.,][]{stanford05,brodwin06,elston06,eisenhardt08,muzzin09, papovich10,brodwin11,santos11,gettings12,stanford12,zeimann12}, but 
exploration  of this high redshift frontier has proven challenging. The challenge 
arises because observable signatures that are commonly used for cluster 
detection (e.g., X-ray and optical flux) diminish toward high redshift, and also 
because massive clusters become increasingly rare earlier in the universe.

Recent years have seen the emergence of a new generation of dedicated surveys that 
identify massive galaxy clusters via the Sunyaev Zel'dovich (SZ) Effect. Several SZ 
galaxy cluster surveys are underway; The Planck satellite \citep{planck13-XX}, the 
Atacama Cosmology Telescope \citep[ACT;][]{marriage11b,hasselfield2013}, and 
the South Pole Telescope \citep{staniszewski09,vanderlinde10,williamson11,reichardt13} 
have all produced SZ galaxy cluster catalogs. SZ Effect 
surveys with sufficient angular resolution to resolve galaxy clusters on the sky 
(e.g., ACT and SPT) benefit from an approximately flat selection in mass beyond 
$z \gtrsim 0.3$ \citep{carlstrom02}, which results in samples with a clean selection 
extending into the $z > 1$ universe. 
From the first 720 (of 2500) deg$^{2}$ of the SPT-SZ survey, 10 clusters have been 
confirmed \citep[regarding the meaning of ``confirmed'' see][]{song12}
at $z > 1$, including six spectroscopically \citep{brodwin10,foley11,stalder13, 
song12,reichardt13,ruel14}. In this work we present spectroscopic observations of 
the highest redshift cluster in the first 720~\sqdeg of the  SPT-SZ survey. \cluster~is the 
most distant galaxy cluster yet discovered via the SZ effect, the second most 
distant cluster with an SZ measurement after IDCS J1426.5$+$3508 
\citep[$z = 1.75$, M$_{200}$ $=$ 4.3 $\pm$ 1.1 $\times$ 10$^{14}$ M$_{\odot}$;][]{brodwin12}, 
and one of only a few spectroscopically confirmed galaxy clusters currently known 
at $z > 1.4$.

This paper is organized as follows. In Section~\ref{s:obs} we describe the 
observations that were critical to the work presented and their reduction. 
In Section~\ref{s:results} we identify spectroscopically confirmed galaxy members 
in \cluster, and report their star formation rates, along with the mass and dynamics of 
the cluster. In Section~\ref{s:discussion} we discuss the properties of the spectroscopic 
cluster members in color-magnitude space, and explore the implications of the high 
incidence of star formation among the cluster members. Finally, we briefly summarize 
our results in Section~\ref{s:conc}.
Throughout this paper we present magnitudes calibrated relative to Vega, 
and calculate cosmological values assuming a standard flat cold dark matter with a 
cosmological constant ($\Lambda$CDM) 
cosmology with $H_{0}=70$ km s$^{-1}$ Mpc$^{-1}$, and matter density 
$\Omega_{M}=0.27$ \citep{komatsu11}.

%%%%%%%%%%%%%%%%%%%%
%%  Observations  %%
%%%%%%%%%%%%%%%%%%%%

\section{Observations and Data}\label{s:obs}

\begin{deluxetable*}{lcccc}[h]
\tablecaption{Imaging Observations of \cluster \label{imagingtable}}
\tablewidth{0pt}
\tabletypesize{\scriptsize}
%\tabletypesize{\tiny}
\tablehead{
\colhead{UT Date}  &
\colhead{Telescope/Instrument} &
\colhead{Filters} &
\colhead{Exp. Time (s)} &
\colhead{Depth\tablenotemark{a}} 
}
\startdata
2010 Oct 29 & CTIO 4m/MOSAIC-II  & $g,r,i$  & 750,1200,1347  & 23.5,22.6,21.6  \\
2011 Jul 14 & CTIO 4m/NEWFIRM  &  $K_{s}$ & 960  & 16.4 \\
2011 Nov 3 & CTIO 4m/MOSAIC-II  &  $z$  & 2400  & 21.2 \\
Cycle 7 & {\it Spitzer/IRAC} & 3.6$\mu m$,4.5$\mu m$ & 800,180 & 20.3,18.8 \\
2012 June 10,11 & Magellan-I/FourStar  &  $J$  & 960 & 20.6 \\
2012 Oct 4 & Magellan-II/MegaCam  & $i$'  & 1800 &  23.8
\enddata
%\tablenotetext{a}{~Coordinates give the center of the spectroscopic observations. All coordinates are J2000.0 epoch.}
\tablenotetext{a}{~10$\sigma$ point source depths.}
\end{deluxetable*}

\subsection{Millimeter Observations by the South Pole Telescope}

The SPT-SZ survey \citep{carlstrom11} finished in November 2011, and covered 
2500 deg$^{2}$ at observing frequencies of 95, 150, and 220 GHz to approximate 
depths of 40, 18, and 70 $\mu$K-arcmin, respectively. Clusters are identified in the 
SPT-SZ survey via the SZ effect, the inverse Compton scattering 
of cosmic microwave background (CMB) photons off of hot intra-cluster gas 
\citep{sunyaev72}. The selection threshold of the SPT-SZ survey is expected to 
fall slightly in mass with increasing redshift, and the resulting cluster sample is predicted 
to be $\sim$100\% complete at $z > 0.3$ for a mass threshold of M$_{500} \gtrsim 
5 \times 10^{14}$ \msun~h$_{70}^{-1}$, and at $z > 1.0$ for a mass threshold of 
M$_{500} \gtrsim 3 \times 10^{14}$ \msun~h$_{70}^{-1}$. Details regarding the survey 
strategy and data analysis are detailed in the previous SPT-SZ survey papers 
\citep{staniszewski09,vanderlinde10,williamson11,reichardt13}.

\cluster\ was initially discovered in the first 720 deg$^{2}$ of the SPT-SZ survey 
and reported in \citet{reichardt13}. It was measured to have a SPT detection 
significance, $\xi$, of 6.28, where $\xi$ is a statistic that reports the strength of the 
detection of the Sunyaev Zel'dovich decrement and scales monotonically with 
mass. The SPT detection is centered at ($\alpha,\delta$) = (20:40:59.23, 
$-$44:51:35.6) (J2000.0), and an image of the filtered SPT map is shown in 
Figure~\ref{f:szopt}. In Section 3.3, we report a new SZ mass estimate based on 
its measured SPT significance and our updated redshift measurement since 
\citet{reichardt13}.

\subsection{Optical and Infrared Imaging}

We obtained $gri$~imaging using the MOSAIC-II imager on the CTIO $4\,\mathrm{m}$
Blanco telescope on UT 29 October 2010 and $z$~imaging on UT 3 November 2011. Both 
nights were clear, with seeing of $\sim$1.35\arcsec~in the October 2010 runs, and 0.68\arcsec~
in the $z$-band data taken in Nov 2011. Total integration times were 750, 1200, 1347, and 2400 
seconds in $g$, $r$, $i$, and $z$, to 10-$\sigma$ point source depths of 23.5, 22.6, 21.6, and 
21.2 magnitudes (Vega) in g, r, i, and z, respectively. The MOSAIC-II data were reduced using the 
PHOTPIPE pipeline \citep{rest05a}, and calibrated photometrically using the stellar locus 
regression technique of \citet{high09}.

We also obtained deep follow-up imaging in $i$' with the Megacam imager \citep{mcleod06} 
on the 6.5-meter Clay Magellan telescope on 2012 October 24. These observations consist 
of 9$\times$200s dithered exposures. The exposures were taken in seeing ranging from 
0.7\arcsec\ to 0.9\arcsec, through variable thin cirrus clouds. The Megacam 
data were reduced at the Harvard-Smithsonian Center for Astrophysics with a 
custom-designed pipeline in addition to standard IRAF/mscred routines. After implementing 
pointing refinements, the nine $i$' exposures were co-added to produce a final mosaic with
an effective FWHM of 0.82\arcsec. We calibrate photometry from the final Megacam $i$' 
mosaic by matching hundreds of well-detected, unsaturated objects that are also detected 
in the MOSAIC-II $i$-band imaging described above; this calibration includes a color term 
that accounts for the different throughput curves of the MOSAIC-II $i$ and Megacam $i$' 
filters.

\begin{figure*}[t]
\begin{center}
\epsscale{1.18}
%\epsscale{1.0}
\plotone{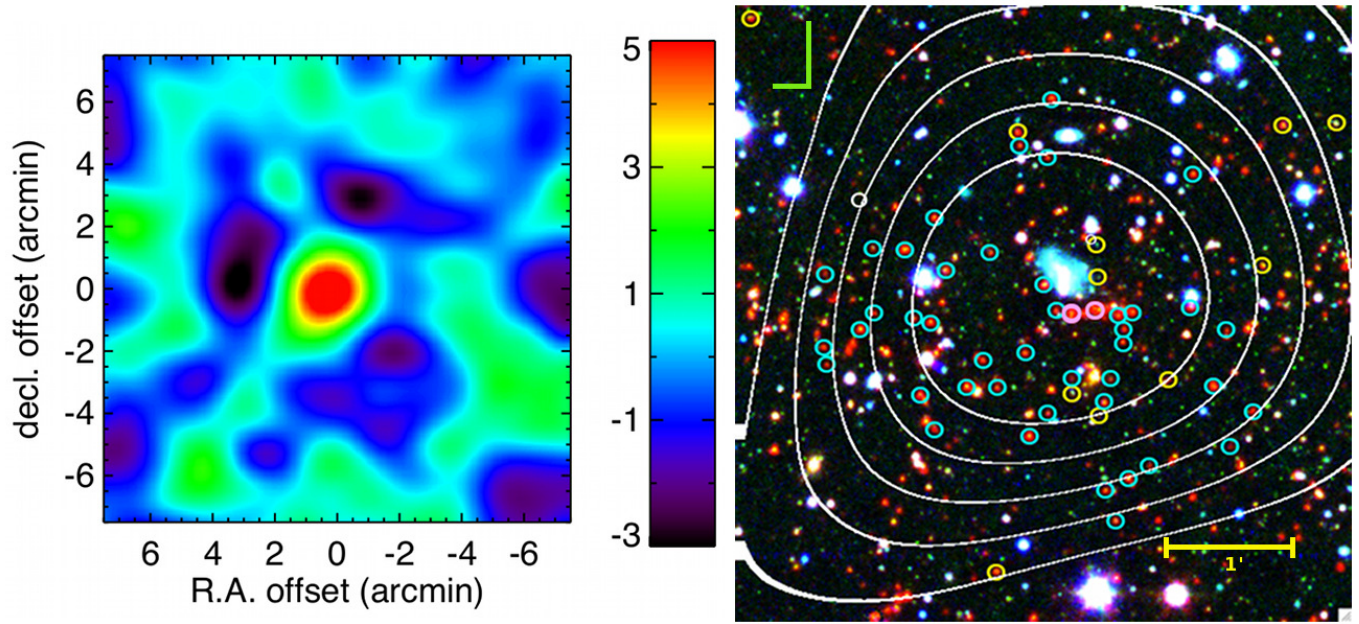}
\caption{
$Left:$ The filtered SPT-SZ significance map of \cluster~with a color map indicating 
significance, $\xi$. The negative trough surrounding the cluster is an artifact of the 
filtering of the time ordered data and maps. $Right$: Color image of the 
4\arcmin$\times$4\arcmin central region around \cluster~from {\it Spitzer}/IRAC [3.6] 
(red) plus Megacam $i$' (green), and MOSAIC-II \emph{r}-band (blue) with the 
SPT-SZ contours over-plotted in white. Photometrically selected cluster members 
are identified with cyan circles, while spectroscopically confirmed cluster members 
are identified with yellow circles. The two candidate brightest cluster galaxies (BCGs) 
are indicated by magenta circles, located near the centroid of the SZ signal. The bright 
blue extended source located near the center of the SZ contours is an intervening 
foreground galaxy. North and East are indicated by the green axes in the upper left 
corner, with North being the longer axis.
}\label{f:szopt}
\end{center}
\end{figure*}

\begin{figure}
\begin{center}
%\epsscale{1.18}
\epsscale{1.19}
\plotone{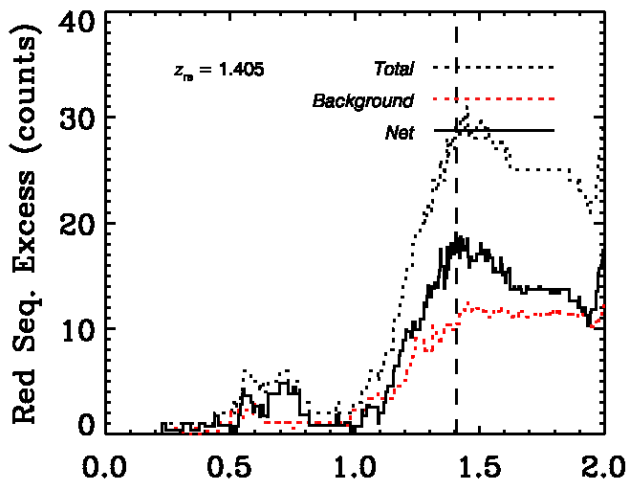}
\caption{
Photometric cluster confirmation, as applied to the SPT cluster sample described in previous SPT 
cluster papers \citep[e.g.,][]{song12,stalder13}. The excess number counts of candidate cluster 
members within a 2\arcmin~radius of the SPT coordinates for \cluster, based on 
\emph{Spitzer}/IRAC 3.6-4.5 colors that are plotted (solid line), along with the total counts 
(dotted black line) and background counts (dotted red line) for comparison. Candidate 
cluster members here are identified as those being consistent within $\pm0.2$ magnitudes of the 
relation expected for a \citet{bruzual03} passively evolving galaxy population that formed at z $=$ 3. 
There is a significant excess of galaxies indicating a cluster at $z_{p} =$ 1.405. 
}\label{f:rsexcess}
\end{center}
\end{figure}

Further ground-based near-infrared imaging was obtained for \cluster~from two different facilities. 
$K_{s}$ imaging with the NEWFIRM imager \citep{autry03} at the CTIO $4\,\mathrm{m}$ 
Blanco telescope was obtained on UT 14 July 2011. Conditions during the observations were 
intermittently cloud with highly variable seeing. The $K_{s}$ observations consist of 
60 second exposures divided among 6 coadds in a 16 point dither pattern, and were reduced 
with the FATBOY pipeline modified to work with NEWFIRM data in support of the Infrared 
Bootes Imaging Survey from the original version developed for the FLAMINGOS-2 instrument 
 \citep{gonzalez2010aas}. SCAMP and SWarp were used to combine individual processed 
frames.  Additionally, $J$-band imaging with Magellan/Baade$+$Fourstar 
was collected on UT 10 \& 11 June 2012 in photometric conditions. A total of 30x32s 
exposures were 
taken at 15 different pointed positions centered on the coordinates of the cluster. The images 
were flat-fielded using standard IRAF routines; WCS registering and stacking were done using 
the PHOTPIPE pipeline. The final $J$ and $K_{s}$ images were calibrated photometrically 
to 2MASS \citep{skrutskie06}, and have FWHM of 0.58\arcsec~and  2.6\arcsec~in $J$ and 
$K_{s}$, respectively.

Infrared imaging for \cluster~was acquired in 2011 with {\sl Spitzer}/IRAC \citep{fazio04} 
as a part of a larger {\sl Spitzer} Cycle 7 effort to follow up clusters identified in the SPT \
survey. The on-target observations consisted of $8\times100$\,s and $6\times30$\,s 
dithered exposures in bands [3.6] and [4.5], reaching 10-$\sigma$ depths of 20.3 and 18.8 
magnitudes, respectively, with an effective spatial FWHM of $\sim$1.66\arcsec. The [3.6] 
observations are sensitive to passively evolving cluster galaxies down to 0.1 $L^*$ at 
$z = 1.5$. The data reduction is identical to that in \citet{brodwin10}, applying the method 
of \citet{ashby09}. All imaging observations are summarized in Table~\ref{t:imagingtable}, 
and we show an IRAC$+$optical+SZ contour image of the core of \cluster~in Figure~\ref{f:szopt}. 
The red sequence excess of galaxies associated with \cluster~in the IRAC imaging data is also 
shown in Figure~\ref{f:rsexcess}. All magnitudes are reported in the Vega system.

\begin{figure*}
\begin{center}
\epsscale{1.175}
%\epsscale{1.0}
\plotone{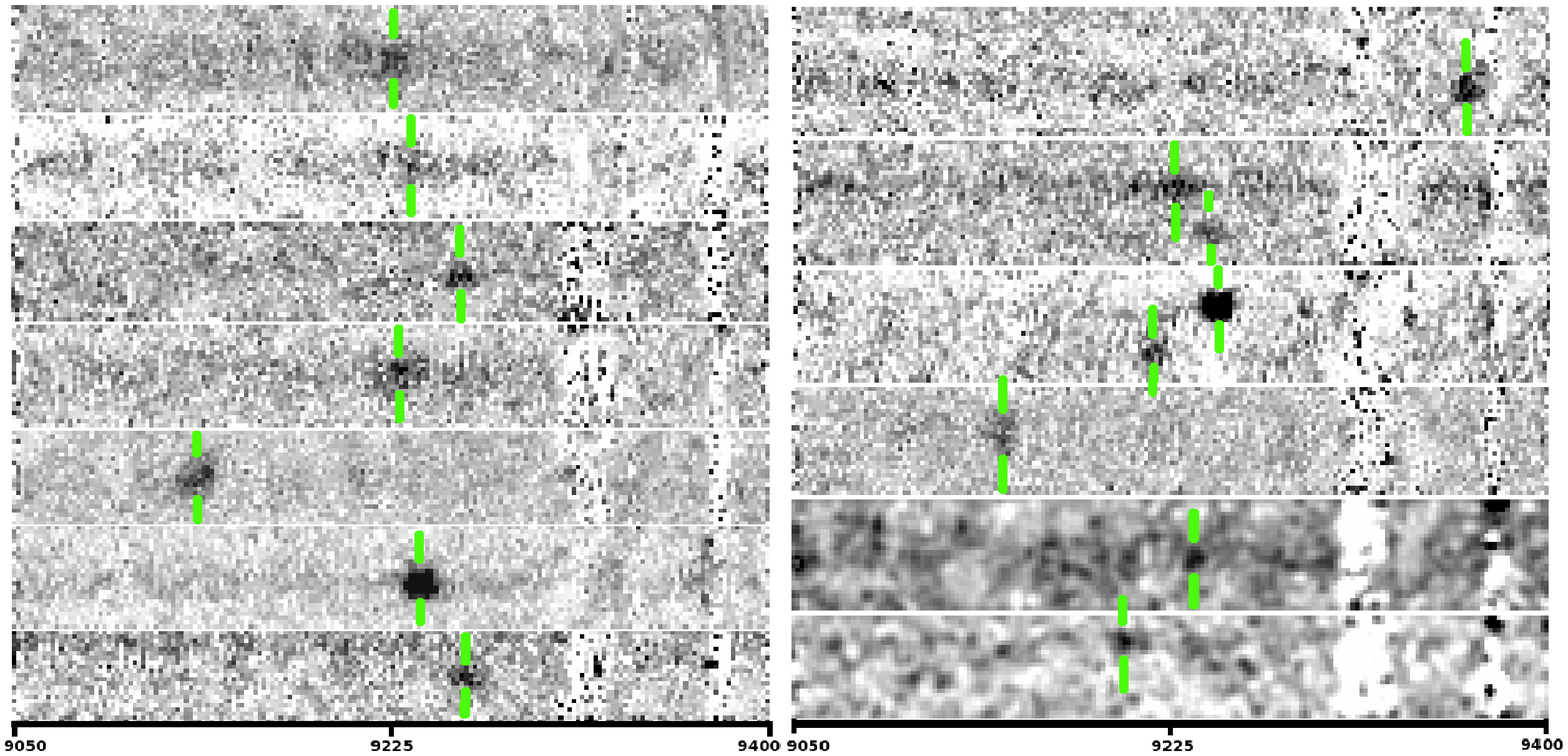}
\caption{
Two dimensional sky-subtracted IMACS spectra containing the 15 [O II] emitting cluster member galaxies, with vertical green brackets indicate the [O II] emission. Each individual 2D spectrum spans the wavelength range, 9050\AA~--9400\AA. The two cutouts in the lower right have each been smoothed with a 2-pixel boxcar kernel to highlight the lower signal-to-noise detections in those spectra. Note that the 2D spectra data in the second and third cutouts from the top on the right side contain the two pairs of galaxies discussed in Section 3.4.
}\label{f:2dspec}
\end{center}
\end{figure*}

\begin{figure*}
\begin{center}
\epsscale{0.22}
%\epsscale{0.18}
\plotone{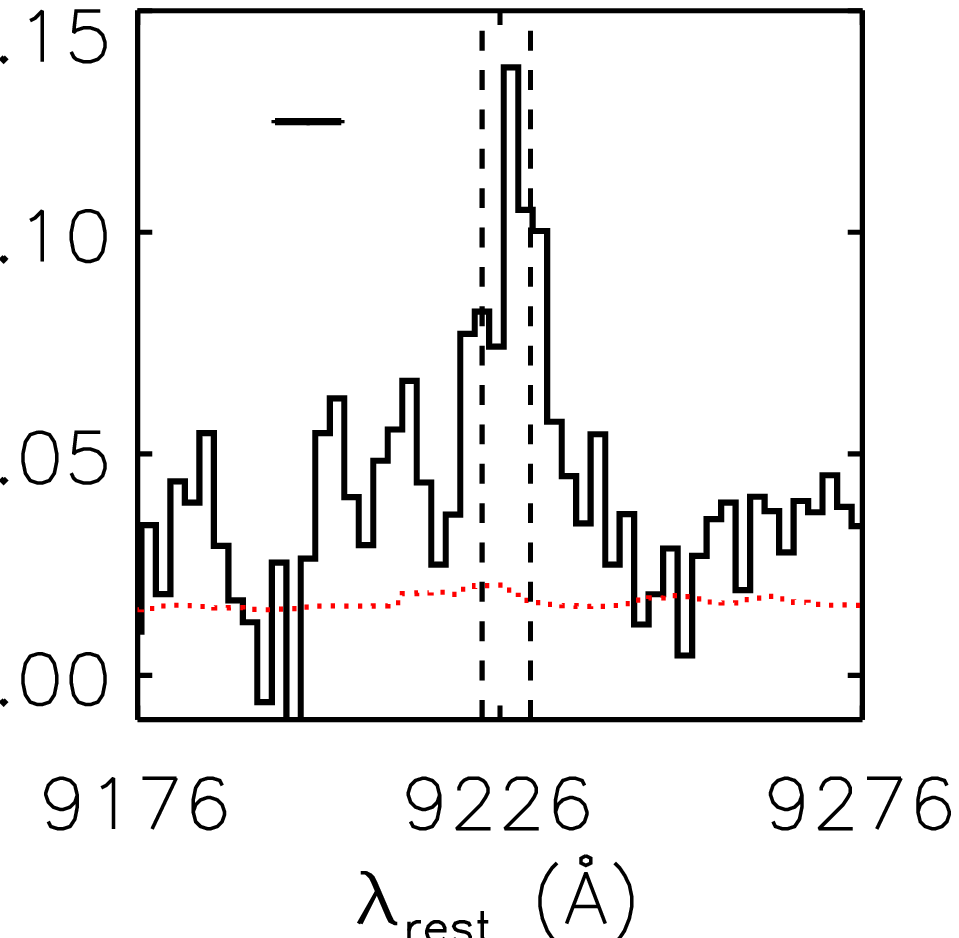}
\plotone{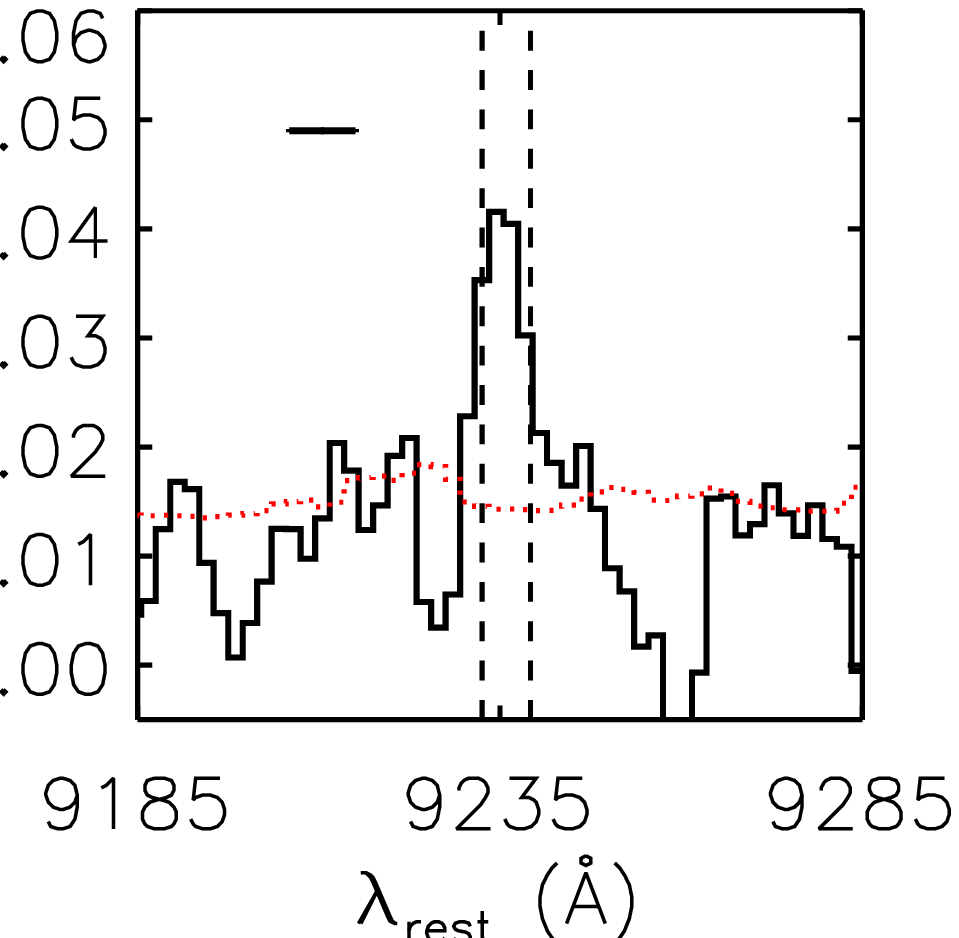}
\plotone{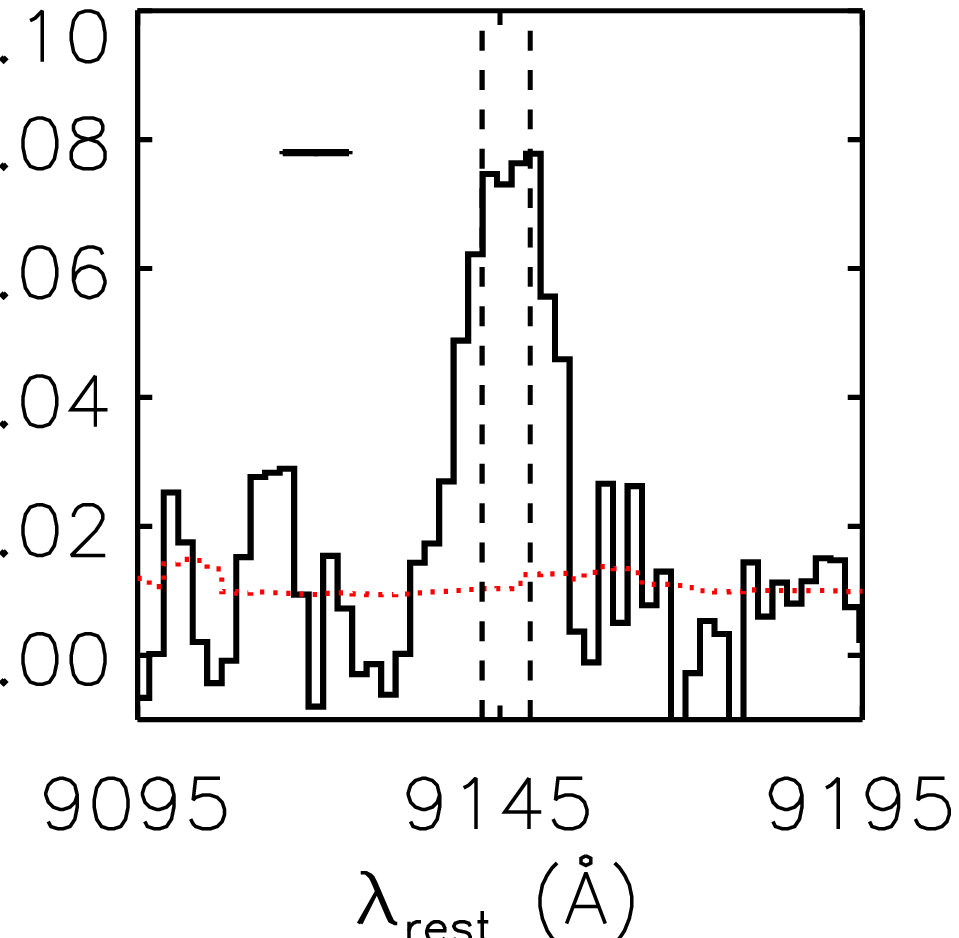}
\plotone{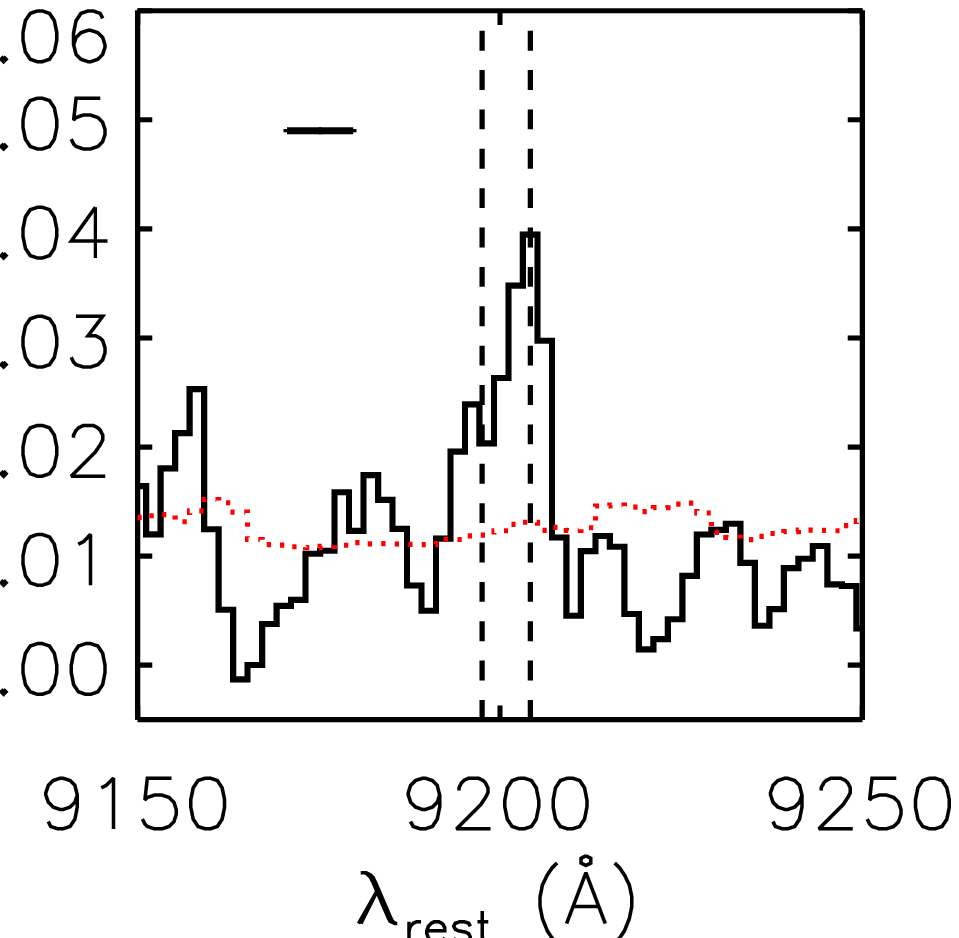}
\plotone{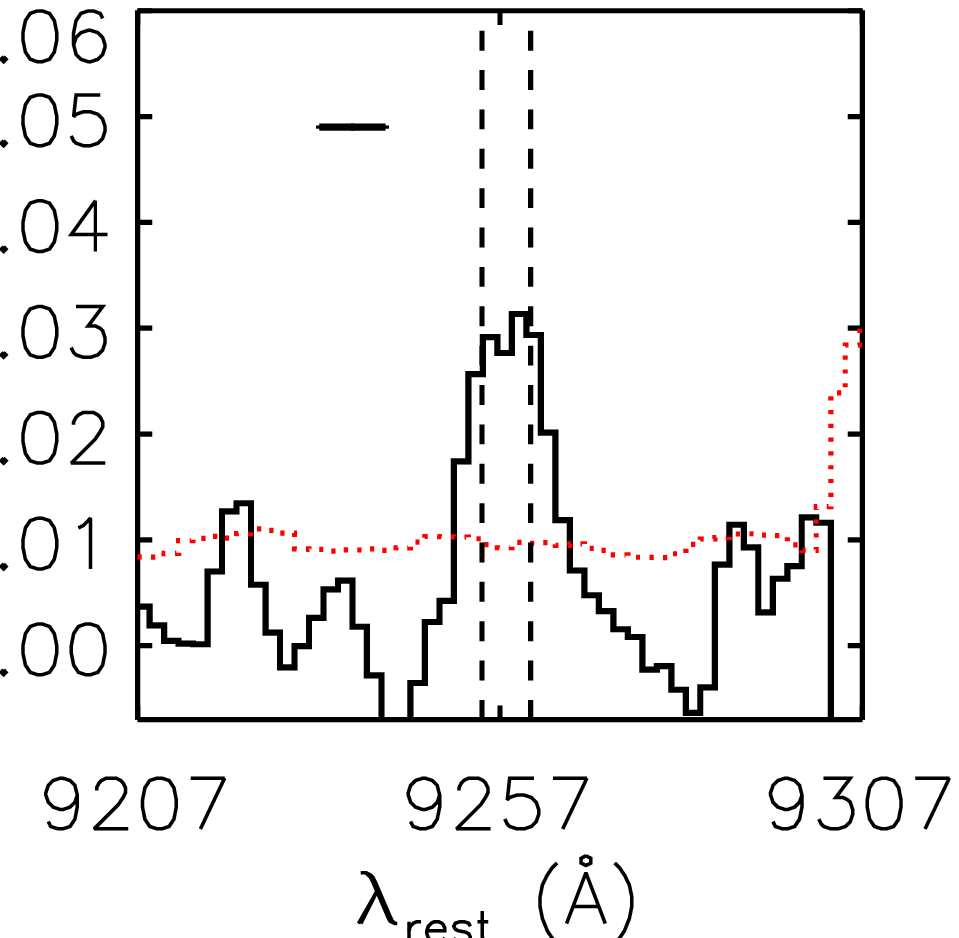}
\plotone{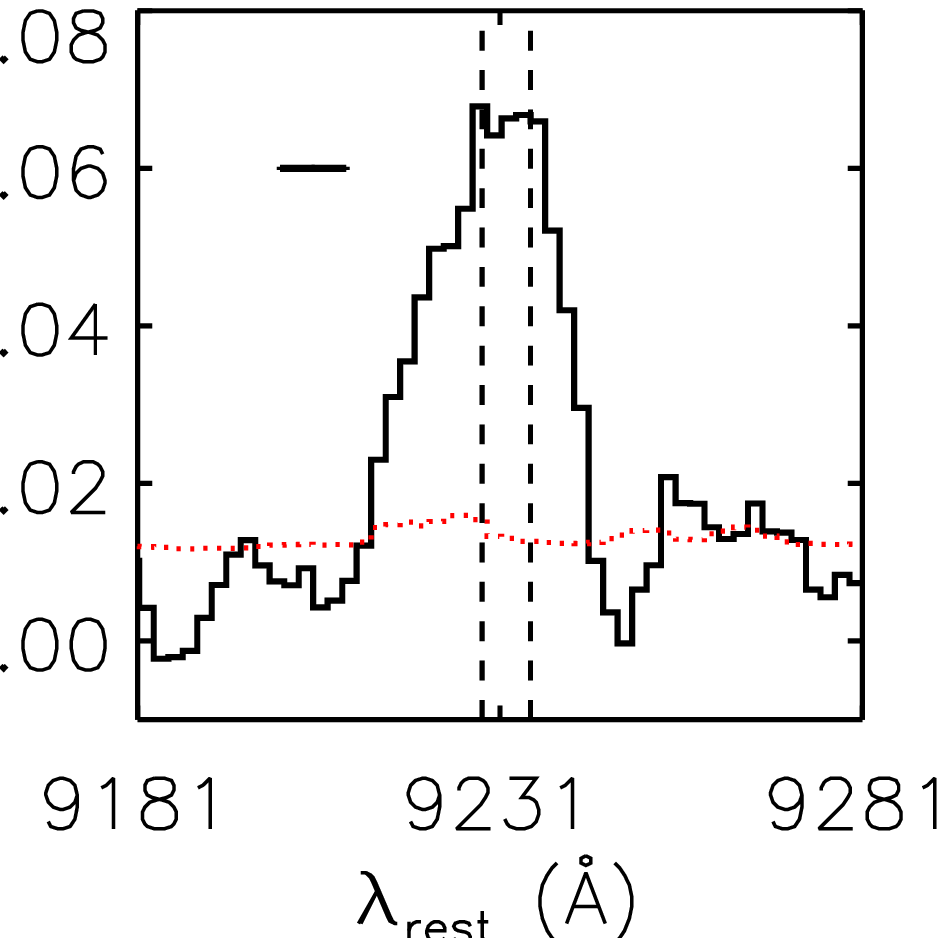}
\plotone{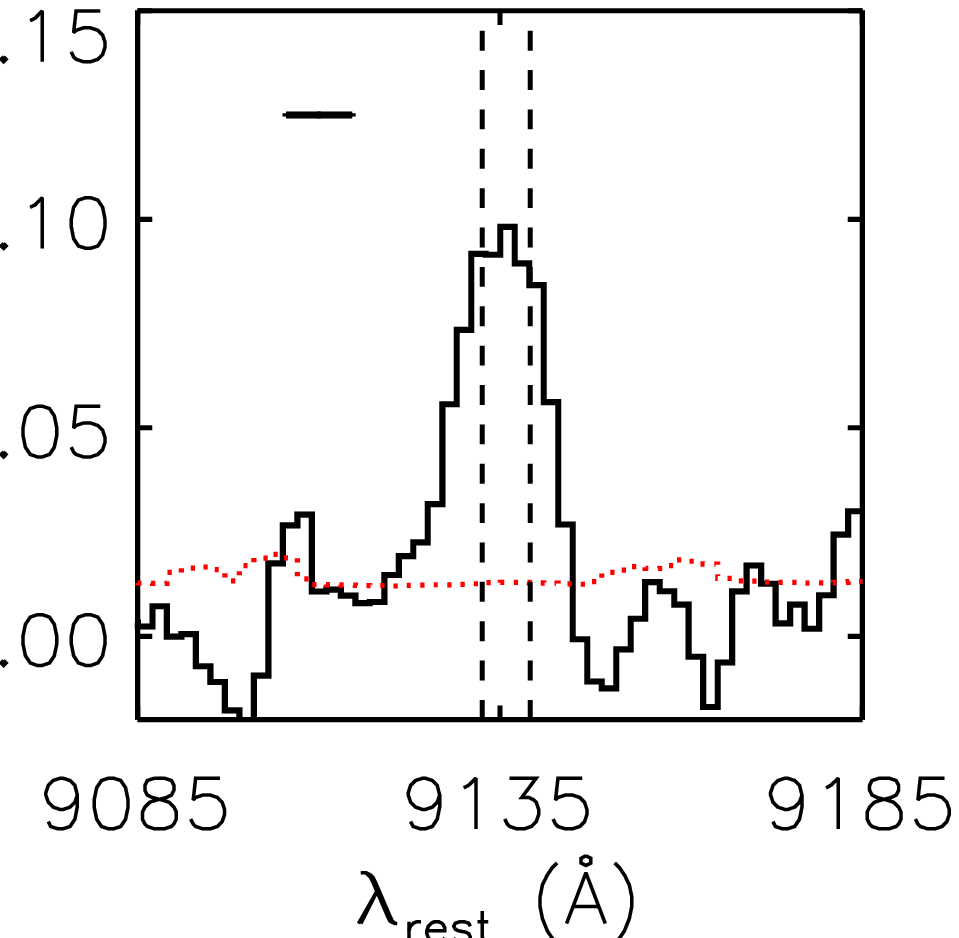}
\plotone{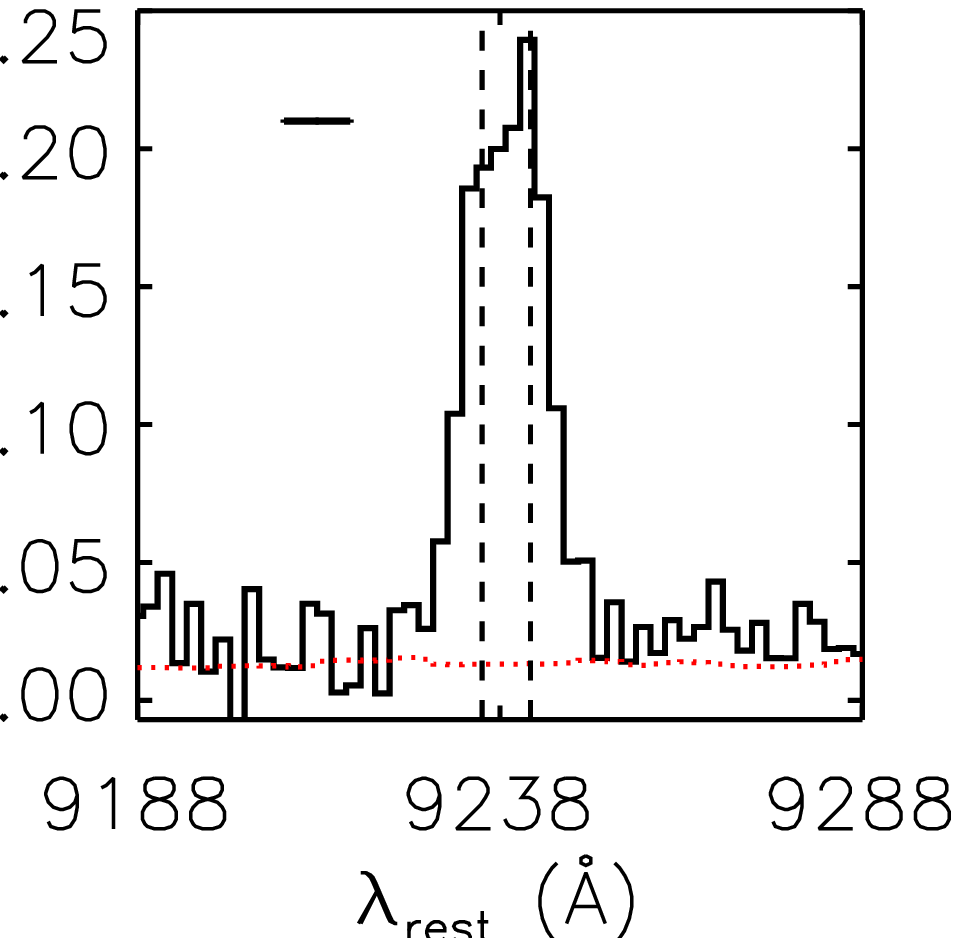}
\plotone{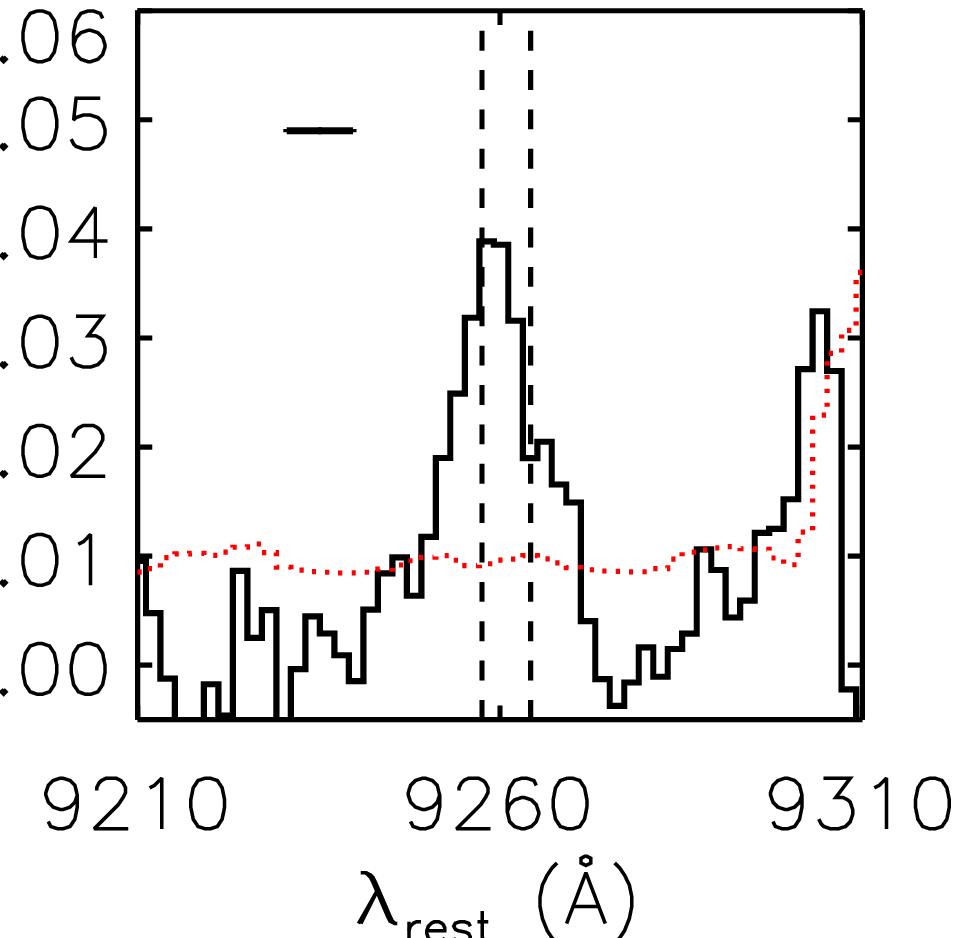}
\plotone{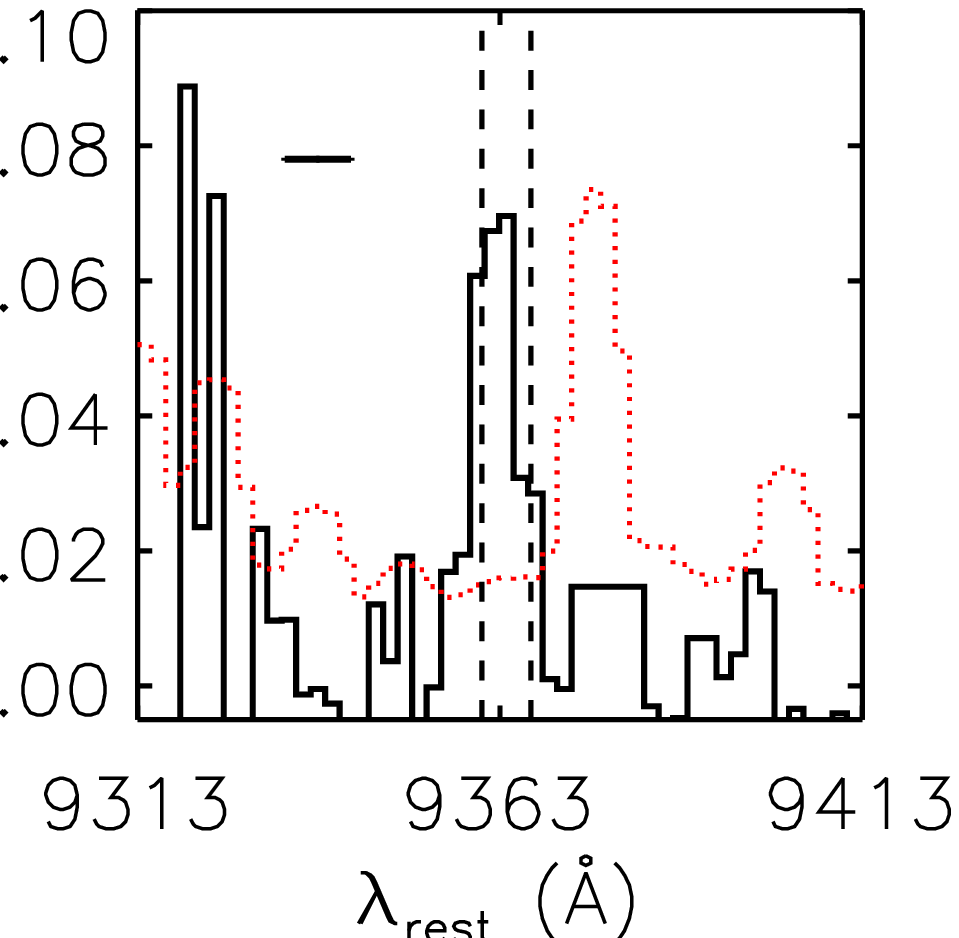}
\plotone{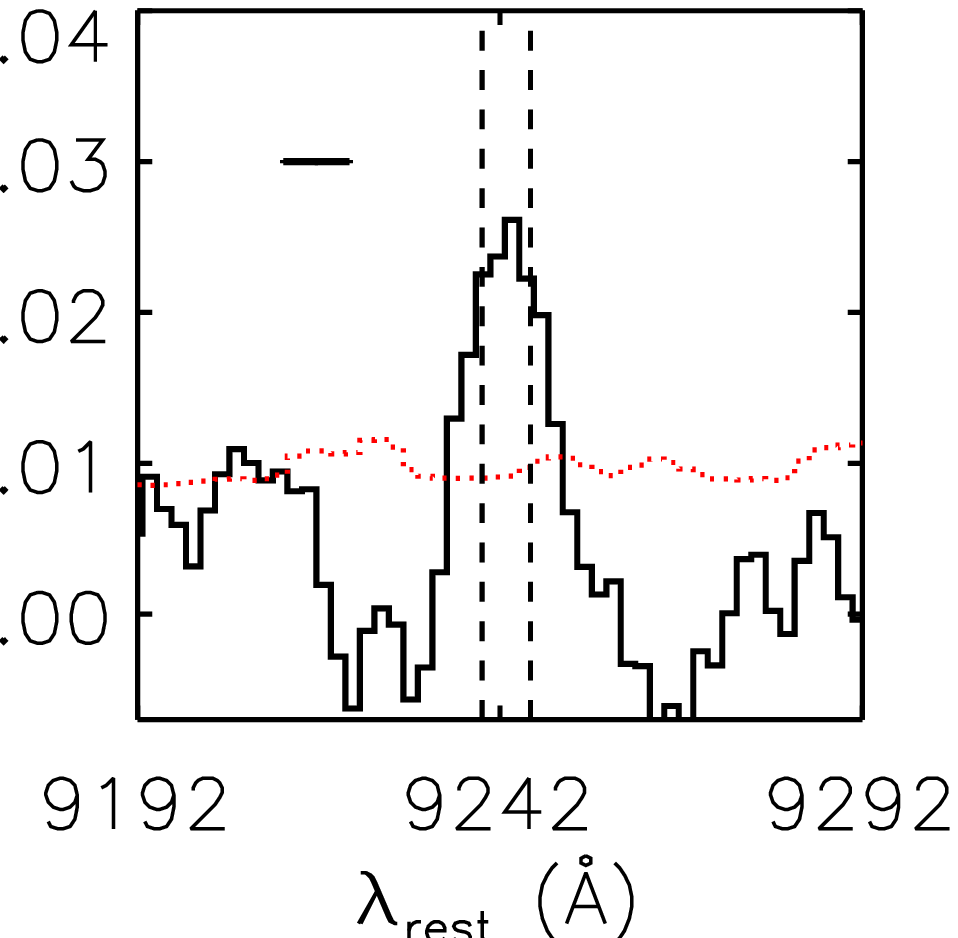}
\plotone{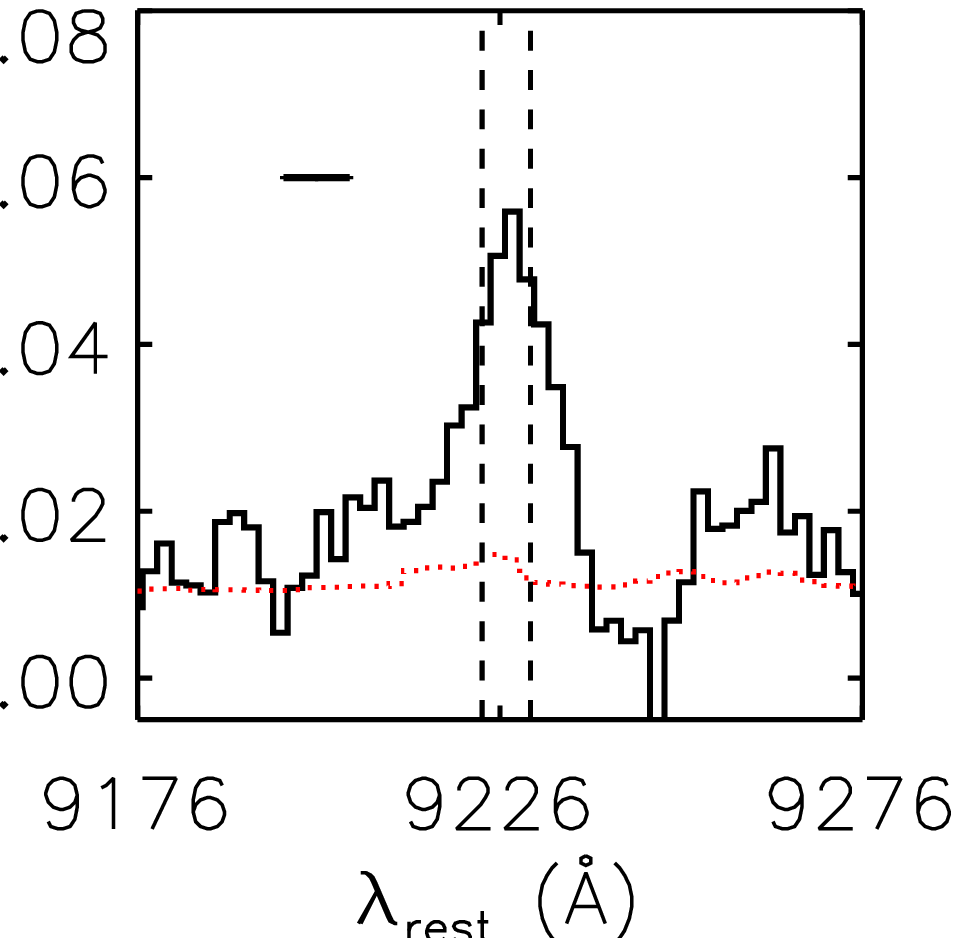}
\plotone{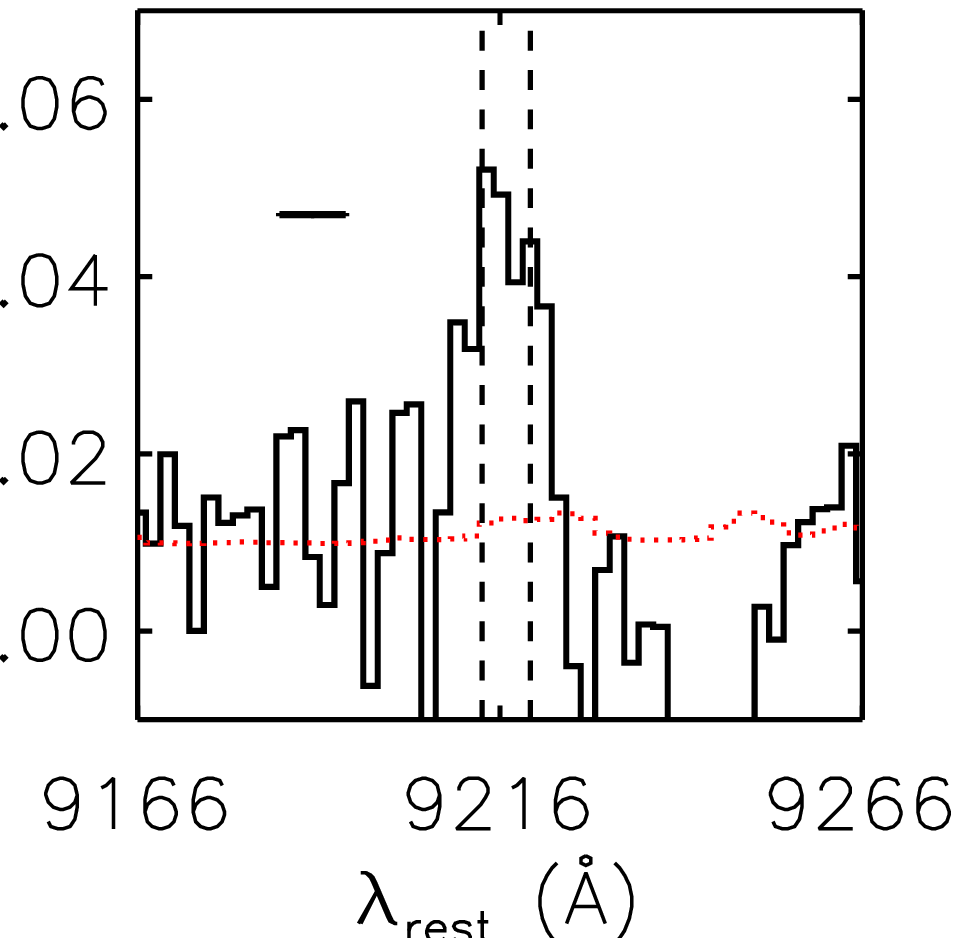}
\plotone{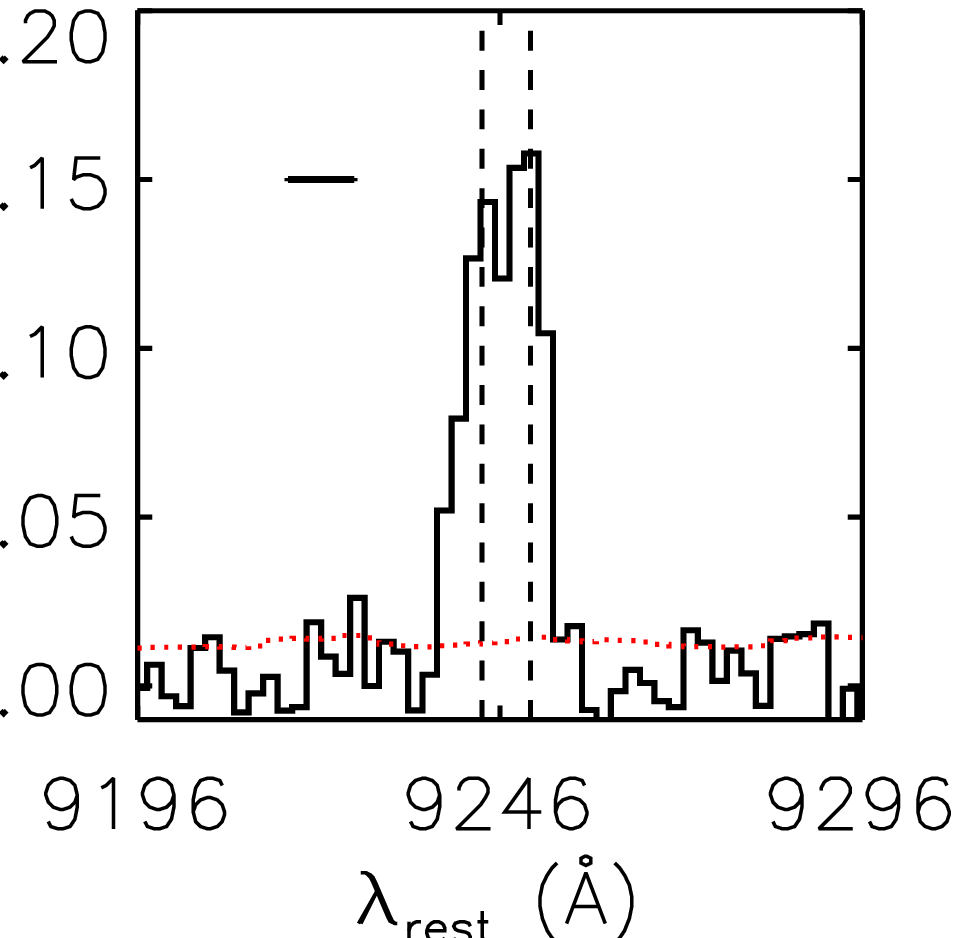}
\plotone{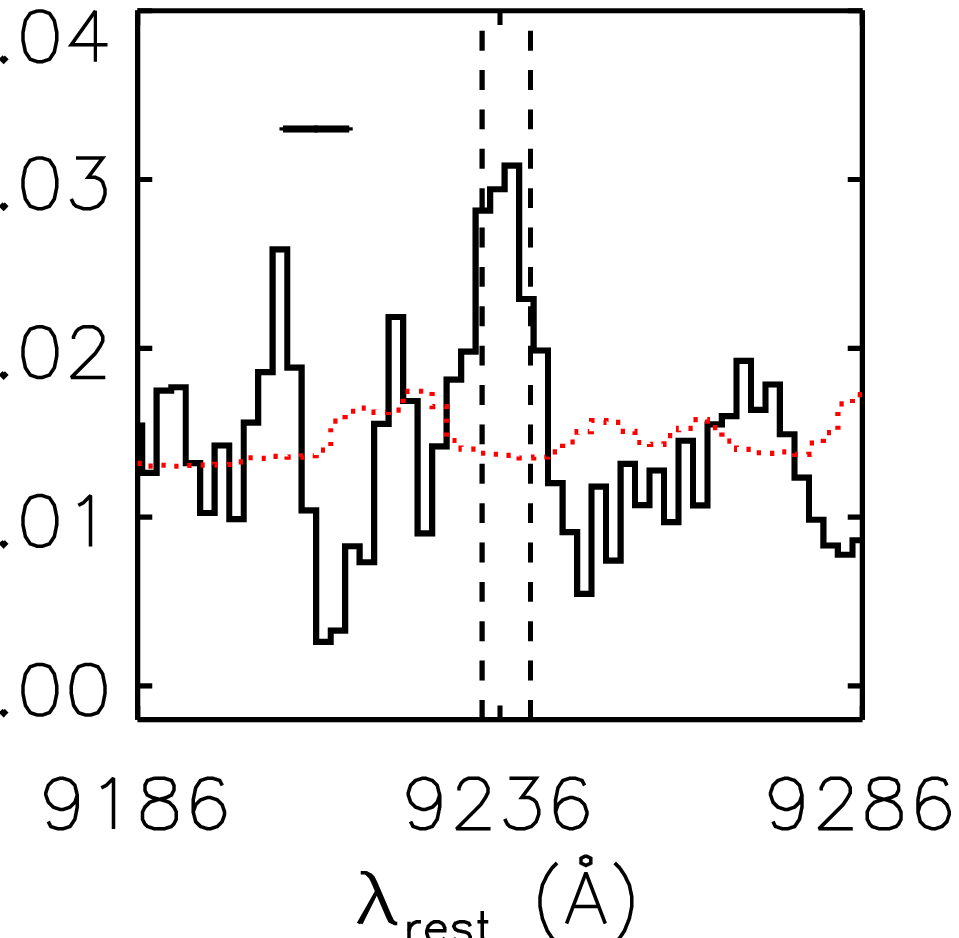}
\caption{ 
Individual 1D spectra of the 15 cluster member galaxies listed in Table~\ref{t:spec} 
spanning an observed wavelength range of $\pm$ 50\AA~ on either side of the 
emission features shown in Figure~\ref{f:2dspec}. The y-axis values in each plotted 
spectrum are in units of erg s$^{-1}$ cm$^{-2}$ \AA$^{-1}$, and the error array for 
each 1D spectrum in the same units is over-plotted as a red dotted line. 
Vertical dashed lines indicate the locations of the redshifted [O II] 
$\lambda\lambda$3727,3729 emission features. Most of the emission features are 
broad or double-peaked, matching the expected [O II] doublet emission profile at 
our spectral resolution (indicated by the horizontal bar in each panel). 
Those lines which are not obviously broad are detected at 
very low signal-to-noise where the morphology of the line is not likely to be well 
measured at all.
}\label{f:1dspec}
\end{center}
\end{figure*}

\subsection{Optical Spectroscopy}\label{ss:specobs}

Spectroscopic observations for \cluster~were carried out on the 6.5-meter 
Baade Magellan telescope on UT 15-16 September 2012 using the f/2 camera on the 
IMACS spectrograph with the 300-line grism at a tilt angle of 26 deg. The f/2 camera 
allows for slits to placed in a circular region with a diameter of $\sim$27'. First night 
observations used the WBP 5694-9819 filter. After measuring a preliminary redshift of $z = 
1.48$ -- somewhat higher than the photometric redshift, $z_{p} =$ 1.41 $\pm$ 0.07 -- for 
numerous galaxies in the first night's data we modified the setup to include no 
spectroscopic filter in order to to be more sensitive to Ca H\&K redward of 
$\sim$9800\AA. The gain in sensitivity due to this change was negligible, as the 
throughput of the IMACS detectors drop of sharply redward of 9800\AA. Spectra of 
individual galaxies cover a typical wavelength range, $\lambda = $ 5700-9820\AA.

The galaxy target selection for mask design was based on the optical and infrared 
photometry presented in \citet{song12}. That analysis identifies 62 candidate cluster 
member galaxies in Spitzer IRAC [3.6]$-$[4.5] vs. [3.6] color-magnitude space.  
There is a strong sequence that forms for galaxies at a common redshift in this 
color-magnitude space \citep[e.g.,][]{brodwin06,muzzin13}, and we use it as our primary 
selection for likely cluster member galaxies. We refined the prioritization by using 
our available optical data to give highest priority to [3.6]$-$[4.5] vs. [3.6] cluster 
candidates with faint counterparts in the $z$-band, and we reject candidates with 
bright counterparts in multiple optical bands (e.g., $i$' $< $ 21); likely low redshift 
interlopers. Two multi-slit masks were designed with 1.2'' wide slits; this slit width 
choice throws away less light from our faint target galaxies, and the loss of spectral 
resolution does not significantly impact our ability to measure redshifts. The first mask 
was observed for a total integration time of 5.7 hrs on UT 15 Sept, and the second mask 
for 5.6 hrs on UT 16 Sept. Both nights were photometric with seeing between 
$\sim$0.6-0.9\arcsec, and using the trace of a point source that fell within one of our 
slits we measure a spatial FWHM (along the slit axis) of 0.85\arcsec\ in our final stacked 
2D spectra.

We use the COSMOS reduction package\footnote{http://code.obs.carnegiescience.edu/cosmos} to 
bias subtract, flat field, wavelength calibrate and sky-subtract the raw data, resulting in 
wavelength-calibrated 2D spectra. The 1D spectra are then boxcar extracted from individual 
source traces in the reduced data. The spectra are flux calibrated from observations of 
spectrophotometric standard LTT 1788 \citep{hamuy1994} taken during the run. Time-series 
of the integrated flux measured for guide stars and DIMM stars throughout the nights of the run 
indicate that the nights were both photometric, with no evidence for significant changes in the 
atmospheric extinction across the two nights of the observing run. We find that the uncertainty in 
the flux calibration is dominated by variable slit losses over the course of the night that result from 
fluctuations in the seeing on timescales of minutes. We measure the scatter in the flux normalization 
directly from our data by measuring the variation in flux measured for well-detected objects in our 
masks across the individual exposures throughout the entire observing run. We find a scatter of 
$\pm$20\%, which results primarily from variations in slit losses over the course of the observations, 
consistent with slit loss variations from changes in the seeing and small variations in the exact 
alignment of the slit-masks on the sky.

Spectral features are identified by eye in the 2D and 1D spectra, and cluster member 
redshifts are measured using the centroid of the blended [O II] line emission (we generally do 
not resolve the individual lines). The FWHM spectral resolution of the observations, as measured 
from sky lines that were extracted and stacked into 1D spectra in the same way as the science 
spectra, is $9.3$\AA. From simulations using the noise properties of our reduced data we find 
that the the final extracted spectra are sensitive to emission line fluxes $>$ 3.8 $\times$ 10$^{-18}$ 
erg cm$^{-2}$ s$^{-1}$ within a spectral resolution element in the wavelength region 
$\lambda \sim$ 9000-9400\AA, which corresponds to the location of [O II]$\lambda\lambda$ 
3727\AA~at the cluster redshift. 

\begin{deluxetable*}{lcccccccccc}
\tabletypesize{\scriptsize}
%  \tabletypesize{\normalsize} 
\tablecaption{Cluster Member Galaxies for
    SPT-CL J2040-4451 \label{t:spec}} \tablewidth{0pt} \tablehead{ \colhead{}
    & \colhead{R.A.} & \colhead{Dec.} & \colhead{}  &
    & \colhead{\scriptsize{Megacam}} &  \colhead{}  
    & \colhead{IRAC} &  \colhead{} 
    & \colhead{IRAC} & \colhead{}  \\ %& \colhead{[O II] Flux} \\
    \colhead{ID} & \colhead{(J2000)} & \colhead{(J2000)} 
    & \colhead{$z$\tablenotemark{a}} & \colhead{$\delta z$\tablenotemark{a}} 
    & \colhead{$i$} &  \colhead{$\delta i$} 
    & \colhead{[3.6]} & \colhead{$\delta$ [3.6]} 
    & \colhead{[4.5]} & \colhead{$\delta$ [4.5]} } \\
%     & \scriptsize{ erg cm$^{-2}$ s$^{-1}$}} \\
    \startdata
    %%%%%%%%%%%%%%%%%%%%%%%%%%%
    %% i Band data has been converted to Vega system
    %%%%%%%%%%%%%%%%%%%%%%%%%%%
J204110.1-444933.6  &  20:41:10.10  &  $-$44:49:33.6  &  1.4760  &  0.0006  &  23.16  &  0.07 & 16.95  &  0.03  &  16.35  &  0.03  \\
J204100.1-445025.2  &  20:41:00.14  &  $-$44:50:25.2  &  1.4777  &  0.0006  &  24.11  &  0.17 &  17.28  &  0.03  &  16.82  &  0.03  \\
J204057.0-445213.7  &  20:40:56.97  &  $-$44:52:13.7  &  1.4842  &  0.0006  &  24.91  &  0.26 &  20.58  &  0.40  &  20.80  &  0.40  \\
J204100.9-445315.7  &  20:41:00.92  &  $-$44:53:15.7  &  1.4765  &  0.0006  &  23.17  &  0.12 & 17.43  &  0.03  &  16.88  &  0.03  \\
J204057.2-445121.4  &  20:40:57.20  &  $-$44:51:21.4  &  1.4540  &  0.0006  &  24.01  &  0.15 & 19.43  &  0.04  &  19.91  &  0.20  \\
J204057.3-445108.6  &  20:40:57.27  &  $-$44:51:08.6  &  1.4693  &  0.0009  &  23.42  &  0.08 &  18.97  &  0.33  &  18.99  &  0.17  \\
J204113.6-445125.2  &  20:41:13.60  &  $-$44:51:25.2  &  1.4509  &  0.0005  &  23.97  &  0.13 &  18.92  &  0.04  &  18.69  &  0.15  \\
J204058.1-445206.7  &  20:40:58.14  &  $-$44:52:06.7  &  1.4789  &  0.0005  &   22.86  &  0.07 & 18.48  &  0.03  &  18.22  &  0.04  \\
J204054.6-445201.1\tablenotemark{b}  &  20:40:54.61  &  $-$44:52:01.1  &  1.4842  &  0.0006  &   23.65  &  0.10 & 18.09  &  0.07  &  17.69 &  0.11  \\
J204051.2-445116.8 &  20:40:51.15  &  $-$44:51:16.8  &  1.5120  &  0.0006  &  22.96  &  0.08 & 17.86  &  0.03  &  16.99  &  0.03 \\
J204050.3-445020.5\tablenotemark{b}  &  20:40:50.27  &  $-$44:50:20.5  &  1.4800  &  0.0006  &  24.60  &  0.21 & 18.98   &  0.25  &  19.20   &  0.40  \\
J204048.5-445021.4  &  20:40:48.52  &  $-$44:50:21.4  &  1.4727  &  0.0006  &  22.06  &  0.06 &  18.09  &  0.03  &  17.49  &  0.03 \\
J204044.2-445124.0  &  20:40:44.24  &  $-$44:51:24.0  &  1.4782  &  0.0006  &  22.91  &  0.07 &   17.34  &  0.03  &  16.77  &  0.03  \\
J204050.4-445022.2  &  20:40:50.42  &  $-$44:50:22.2  &  1.4758  &  0.0006  &  23.01  &  0.08 & 17.22  &  0.03  &  16.69  &  0.03  \\
J204048.7-445020.7  &  20:40:48.73  &  $-$44:50:20.7  &  1.4808  &  0.0006  &  23.32  &  0.08 &  19.15  &  0.14  &  18.84  &  0.10 
\enddata
\tablenotetext{a}{Spectroscopic redshifts are measured from the blended [O II] doublet line emission, and uncertainties are dominated by the uncertainty in the centroid of a profile fit to the emission.}
\tablenotetext{b}{These objects are adjacent to and blended with other bright sources in the IRAC imaging. We mask/subtract the contaminating light from these sources to measure the IRAC fluxes reported here, 
and assign them large errors reflecting the systematic uncertainty in the masking/subtraction.}
\end{deluxetable*}

%%%%%%%%%%%%%%%
%%  Results  %%
%%%%%%%%%%%%%%%

\section{Results}\label{s:results}

\subsection{Cluster Member Galaxies}\label{ss:members}

\cluster~was initially measured to have a photometric redshift of \zp~by fitting a model of 
passively-evolved galaxies from \citet{bruzual03} to the available optical$+$NIR 
data; this process is described extensively in \citet{song12}. Incorporating additional 
follow-up data -- specifically the Fourstar $J$-band and Megacam $i$-band -- refines 
the photometric redshift measurement to $z_{p} = $ 1.40 $\pm$ 0.06. At this redshift, 
we expected our IMACS observations to be sensitive to numerous spectroscopic 
features in cluster member spectra, including [O II]$\lambda$$\lambda$ 3727, CaII 
H\&K, and the 4000\AA~break. The IMACS spectra resulted in 15 galaxies with clear 
emission lines visible in the reduced 2D spectra in the wavelength range 9140\AA~$<$ 
$\lambda_{obs}$ $<$ 9370\AA~(Figure \ref{f:2dspec}), and no other emission lines 
elsewhere along the entire spectral trace extending to the blue limit of the spectra 
($\sim$5800\AA). Spectroscopic and photometric measurements of these likely cluster 
member galaxies are summarized in Table~\ref{t:spec}.

\begin{figure}
\begin{center}
\epsscale{1.2}
%\epsscale{0.95}
\plotone{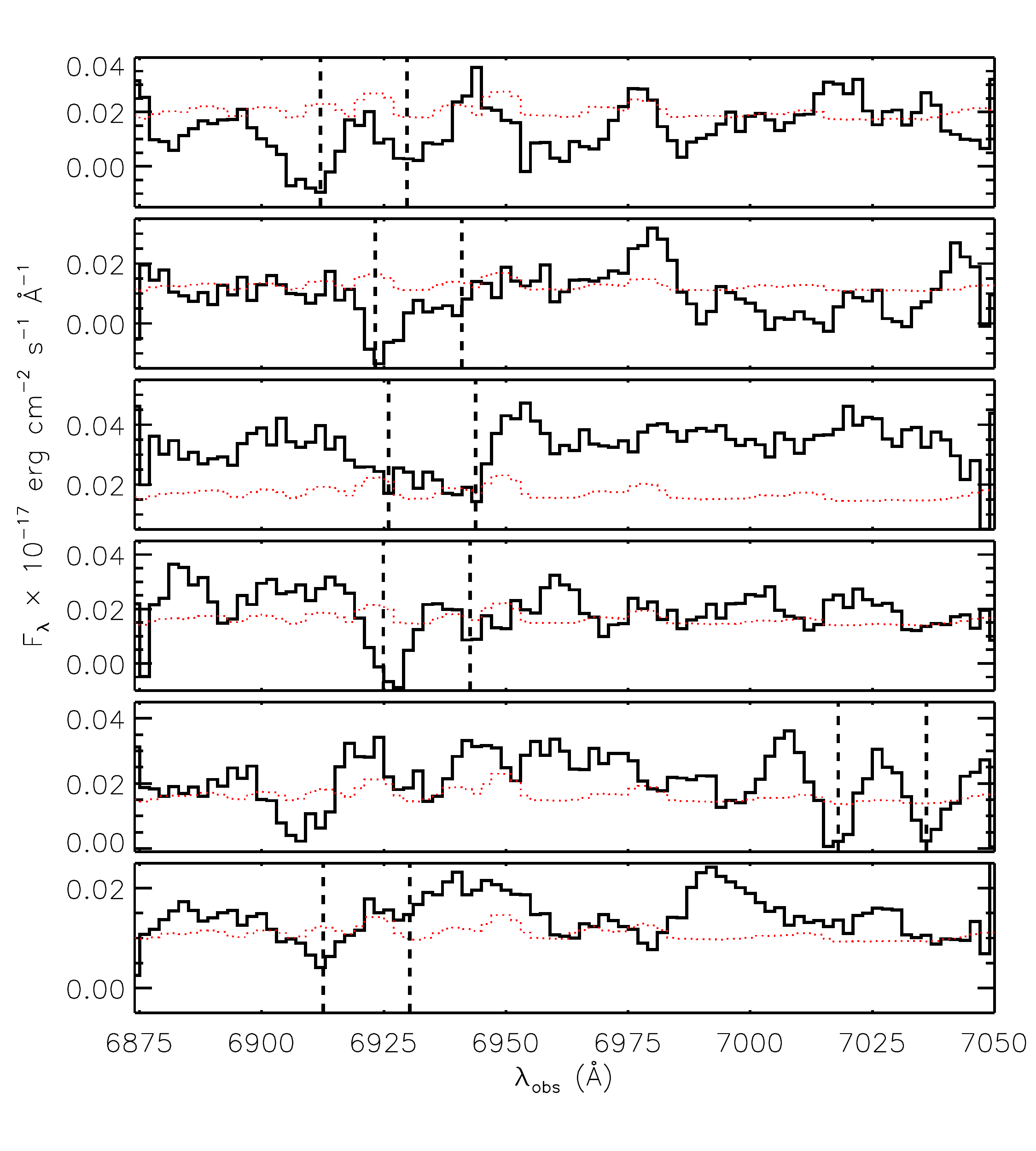}
\caption{
Extracted 1D spectra in the wavelength interval 6875\AA~$<$ $\lambda_{obs}$ $<$ 
7050\AA~ for the six spectroscopic cluster members that exhibit low S/N absorption 
features that are consistent with MgII$\lambda\lambda$2796,2803. The error array for 
each spectrum is plotted as a red dotted line. The spectra have 
been smoothed by a kernel matching the spectra resolution of the data. Vertical dashed 
lines indicate the locations of the apparent MgII absorption lines. These absorption 
lines are not themselves especially robust, but taken in conjunction with the 
clear emission lines (Figure~\ref{f:2dspec}) confirm $z \simeq 1.48$ for the 
emission line galaxies.}
\label{f:mgii}
\end{center}
\end{figure}

These emission lines are consistent with [O II]$\lambda$$\lambda$ 3727 redshifted to 
$z \sim 1.48$. Furthermore, those lines with large signal-to-noise (S/N) have line widths that 
are broader than the spectral resolution of the observations, consistent with the blended 
profile of the redshifted [O II]$\lambda$$\lambda$ 3727 doublet (e.g., Figure~\ref{f:1dspec}). 
The lack of additional 
emission features blueward of the detected lines supports the hypothesis that these 
features correspond to [O II]$\lambda$$\lambda$ 3727, as the spectral coverage would 
include other bright nebular emission lines if the features that we observe were actually 
H-$\alpha$, H-$\beta$, or O[III]$\lambda$$\lambda$ 4960,5008. Furthermore, six of 
the brightest [O II] emitting galaxies also have weak continuum absorption features that 
match the MgII $\lambda$$\lambda$2796,2803 doublet at the same approximate 
redshift as the [O II] emission features (Figure~\ref{f:mgii}) -- these absorption features 
are blue-shifted with velocities ranging from $-$20 to $-$970 km s$^{-1}$ relative to 
the emission lines in the corresponding spectra (Table~\ref{t:mgii}), as would be 
expected for MgII absorption lines from outflowing gas.

Five of the six outflow 
signatures have v$_{outflow}$ $\lesssim$ 500 km s$^{-1}$, as is typical of outflows 
in the interstellar medium due to winds in star forming galaxies \citep{shapley03}, 
and one has a velocity, v$_{outflow}$ $=$ 970 km s$^{-1}$, similar to those observed 
in the most vigorously star forming galaxies \citep{weiner09}. These MgII features are 
similar to those seen by, e.g., \citet{papovich10} in a galaxy cluster at $z = 1.62$. We 
note that one of the brighter line-emitting galaxies cannot be tested for the presence 
of MgII absorption at $z \sim 1.48$ because the relevant part of the spectrum falls into 
an IMACS chip gap (the IMACS f/2 configuration uses fixed grism dispersers which 
cannot be adjusted to dither spectra along the dispersion direction). Another one of 
the brighter candidate cluster members exhibits possible MgII absorption features 
that are unfortunately coincident in wavelength with the telluric B band, and is 
therefore excluded from Figure~\ref{f:mgii} and Table~\ref{t:mgii}. 

We also make a composite stack of all 15 spectra that we identify as cluster 
members. To stack we shift each spectrum into the rest frame based on the 
[O II]$\lambda\lambda$3727,3729 emission feature, and mapping the shifted spectra 
to a common wavelength array (i.e., flux uniformly binned in wavelength) by linearly 
interpolating the shifted spectra. We then sum the flux from each of the member spectra, 
to produce the stack (Figure~\ref{f:stack}). We explored more complex stacking methods, 
such as median and averaging after applying a variety of sigma-clipping algorithms, but 
the resulting stack is qualitatively insensitive to method (i.e., they all have the same ISM 
absorption features and lack of Ne V emission lines. In this stacked spectrum we identify 
absorption features that correspond to FeII$\lambda\lambda$ 
2586,2800 and MgII $\lambda$$\lambda$2796,2803 at a mean outflow velocity of 
$\sim$120 \kms, consistent with the handful of individual outflow signatures described 
above. In the stacked spectrum we also note a distinct lack of emission corresponding 
to the high-ionization [Ne V] $\lambda$$\lambda$3346,3427, which argues against 
AGN activity as a dominant source of the observed [O II] emission. It is also apparent 
from Figure~\ref{f:stack} that our data are not sufficiently sensitive in the rest-frame 
wavelength range containing the CaII H \& K absorption doublet to allow for a 
detection of those features. 
Based on all of the above evidence we confidently conclude that the 15 observed 
emission lines are [O II]$\lambda$$\lambda$ 3727 from member galaxies in \cluster.

At the spectroscopic redshift of the cluster the spectral features that are typically 
used to identify passive galaxies -- primarily Ca H\&K, and the 4000\AA~break -- 
are redshifted to wavelengths where the instrumental throughput of IMACS is falling 
rapidly toward zero and there are numerous bright sky lines (e.g., Figure~\ref{f:stack}). 
As a result we are unable to measure absorption line redshifts of passive cluster 
members in \cluster~with high confidence in the IMACS data. There are two 
red-sequence galaxies that could be considered the ``brightest 
cluster galaxy'' (BCG), with m$_{3.6\mu m}$ = 16.04 and 16.16. Both of these galaxies 
are a factor of $\gtrsim$2 brighter than the next brightest galaxies at 3.6$\mu m$, 
suggesting that they are substantially higher stellar mass. The presence of two nearly 
equally-bright BCG candidates is reminiscent of the Coma cluster, which is in the late 
stages of a galaxy cluster merger \citep[e.g.,][]{colless96,biviano96}. 
Alternatively, we may simply be observing an epoch at which the dominant galaxy 
had yet to be established. For comparison, \citet{delucia07a} simulate the 
hierarchical formation of BCGs and show that the dominant cluster galaxy may not 
be established until $z \lesssim 1.1$. Unfortunately we did not place a spectroscopic slit 
on the second brightest of these objects, and with the data presented in this work, we are 
unable to differentiate between these two scenarios.

We also note that the spectrum of one of the two potential BCGs -- as described 
above -- had a slit placed on it in the first of our two masks, and the resulting spectrum 
shows clear continuum emission redward of $\sim$8000\AA~with no significant 
emission features, as would be expected for a passively evolving galaxy at $z \sim 1.48$. 
The spectrum is low S/N ($\sim$2 per 
spectral pixel) and shows no strong features, which is typical of passive galaxy spectra 
in the rest-frame wavelength range $\sim$3200-3900 -- i.e., the rest-frame wavelengths 
sampled by our observations redward of 8000\AA~at the cluster redshift. In addition to 
the potential BCGs, several other Spitzer-selected candidate cluster members have 
spectra that exhibit no signal (or S/N well below 1 per pixel) continuum redward of 
8500-9000\AA, also consistent with potentially being passive galaxies at the cluster 
redshift. This consistency does, not, of course, preclude the possibility that some of 
these galaxies are late-type dwarf stars or other interlopers.

\subsection{Star Formation in the Cluster}

\begin{figure*}[t]
\begin{center}
%\epsscale{1.}
\epsscale{1.2}
\plotone{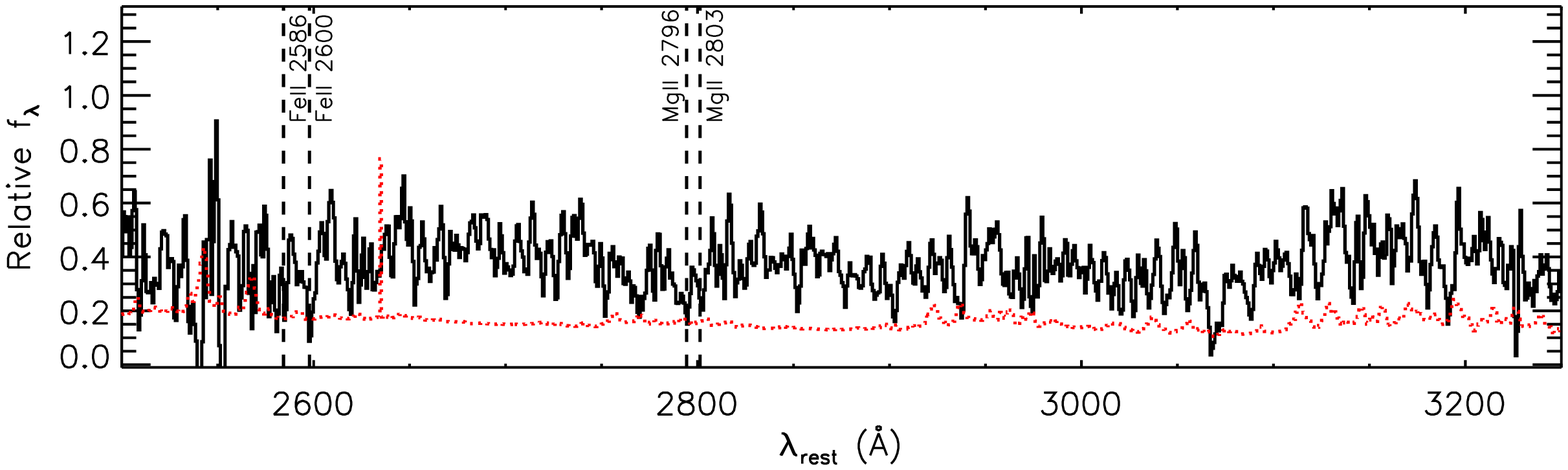}
\plotone{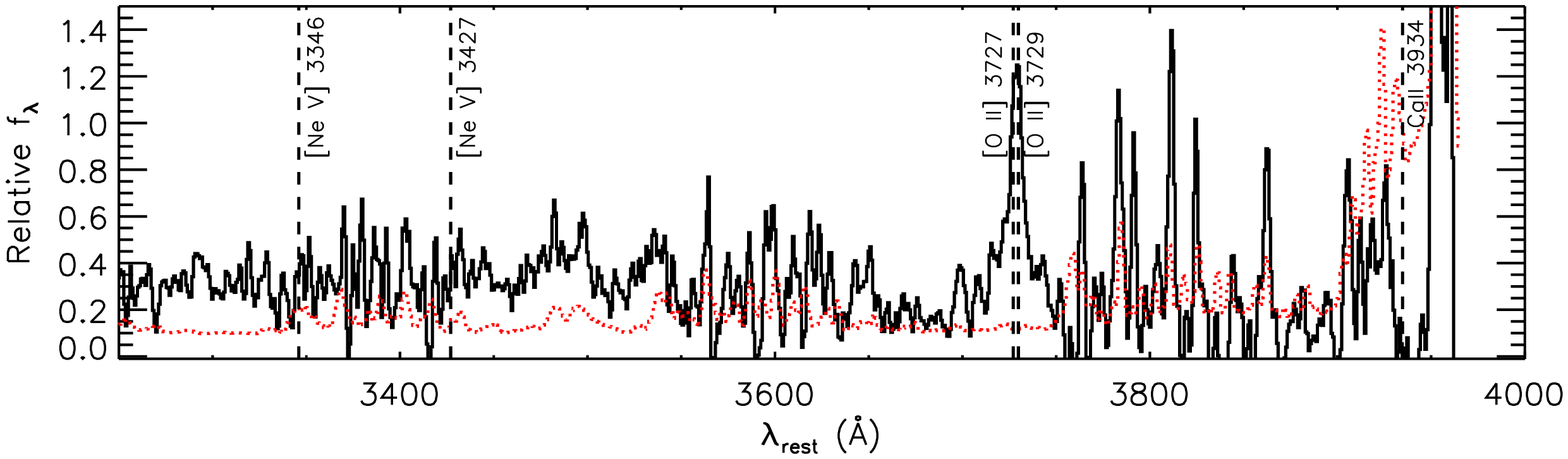}
\caption{ 
A stack of all 15 spectroscopic cluster member spectra, spanning the full 
wavelength range covered by the observations ($\sim$6000-9800\AA~, 
$\sim2500-4000\AA$~in the rest frame). $Top:$ The stacked spectrum in 
the rest-frame wavelength range $\lambda_{rest} = $2500-3250\AA. 
The error array is over-plotted as the red dotted line. Vertical dashed lines 
indicate the presence of strong ISM absorption lines blue shifted by $\sim$ 
120 \kms relative to the [O II]$\lambda\lambda$3727,3729 emission. 
$Bottom:$ The stacked spectrum in the rest-frame wavelength range 
$\lambda_{rest} = $3250-4000\AA, with the error array again 
over-plotted as the red dotted line. Vertical dashed lines here indicate 
the location of [O II]$\lambda\lambda$3727,3729 emission, along with the 
predicted location of several un-detected features: two high-ionization [Ne V] 
forbidden lines (common in AGN) and the CaII K absorption line.
}\label{f:stack}
\end{center}
\end{figure*}

Our IMACS spectra provide [O II]$\lambda$$\lambda$3727 flux measurements or 
lower limits for all 15 spectroscopically confirmed cluster member galaxies. We 
measure the flux by fitting a gaussian to each emission line and and integrating the 
total flux of the gaussian fit. We allow for a local continuum level underneath each 
gaussian fit and subtract the continuum off before integrating; in practice the 
continuum levels are consistent with zero and dwarfed by the emission line flux in 
all cases. The presence of [O II]$\lambda$$\lambda$3727 emission is strong evidence of 
ongoing star formation, but converting from [O II] flux to star formation rate (SFR) is an 
uncertain process \citep[e.g.,][]{yan06,lemaux10}. The observed [O II] line 
luminosity is very sensitive to dust extinction in the rest-frame, but it is also possible for 
the observed [O II] emission to originate from an active galactic nucleus (AGN) rather 
than star formation. AGN emission should be spatially unresolved in our observations, as it 
would originate from a very small physical region in the cores of the galaxies, whereas 
line emission from star forming regions should be distributed throughout the galaxies 
and result in extended emission. As previously noted, we do not find any [Ne V] emission 
the stacked spectrum of the 15 cluster members, which argues against the kind of hard ionizing 
spectrum that would result from strong AGN activity (Figure~\ref{f:stack}). We also find that the 
emission line profiles along the spatial axis (i.e., along the slit) are extended relative to a point 
source (Section~\ref{ss:specobs}) for all but one of the 15 spectroscopic cluster members, 
and this single exception (J204057.0-445213.7) is one of the lower S/N detections in our 
spectroscopic data, where the spatial FWHM measurement is significantly uncertain. 
From the above evidence we conclude that the [O II] that we observe is not likely 
to be AGN-dominated. We cannot rule out a low, sub-dominant level of AGN contribution 
to the measured [O II] fluxes for the member galaxies of \cluster. We do note that it is possible 
that some of the [O II] emission that we observe is associated with Low Ionization Nuclear 
Emission-line Region (LINER) processes. LINER line emission is not directly associated with 
star formation and is sometimes observed to be spatially extended, but is also not 
necessarily associated with AGN activity in all cases \citep{yan12}. From our data we 
lack the information necessary to precisely identify LINER-like galaxies in our sample.

Given [O II]$\lambda$$\lambda$3727 flux measurements we can make a very rough 
attempt at estimate the SFR within each [O II] emitting galaxy. These estimates are, 
however, subject to serious caveats due to corrections that must be made to account 
for slit losses in our spectroscopy, as well as suppressed [O II] emission due to dust 
extinction. \citet{rosa-gonzalez02} provide an empirical prescription for SFR estimates 
based on rest-frame optical and UV observables that attempts to use correlations between 
SFR and dust properties to correct for underestimates of the SFR due to extinction. Using 
[O II]$\lambda$$\lambda$3727 luminosity, this amounts to a factor of 6$\times$ increase 
in the estimated SFRs relative to the \citet{kennicutt98} Case B relation. This is the best 
estimate that we can use to correct for the dust extinction in our galaxies, due to the lack 
of a means to measure the dust extinction within our individual galaxies. Correcting for slit 
losses is a highly uncertain process for the spectra reported in this paper that correspond 
to the serendipitously detected galaxies (those galaxies that fell partially onto a slit that 
was centered on a different source). This is because the serendipitous sources are 
not centered on a slit, and therefore are subject to huge slit loss uncertainties as a function 
of variations in the seeing. For these galaxies we report only lower limits on the total line flux 
due to the extreme uncertainty in computing slit loss corrections; these limits correspond to lower 
limits in the inferred SFR from [O II]$\lambda$$\lambda$3727.

Using a standard cosmology (see Section 1) we 
compute the corresponding luminosity in [O II] along with the corresponding SFR 
assuming Case B recombination and the \citet{kennicutt98} relationships between 
nebular line emission and the rate of star formation. It would be ideal to have additional 
star formation indicators for \cluster, but the available data -- including WISE photometry 
-- are to shallow by more than an order of magnitude to make a measurement or place 
interesting limits. We also measure the projected distance between each cluster member 
and the cluster centroid as measured via the SZ effect; these impact parameters can be 
compared to the radius, R$_{200}$, at which the interior mean density of the cluster 
is 200 times the mean density of the universe at the cluster redshift, $\rho_{m}(z)$. 
The M$_{200,SZ}$ value computed in section 3.3 implies R$_{200}$ for \cluster~of 
1.2 Mpc.  All 15 spectroscopically confirmed cluster members are within 2.85\arcmin 
-- a projected distance of approximately 1.5 Mpc -- of the centroid of the SZ signal as 
measured by the SPT. Individual projected distances, [O II] fluxes, and [O II]-based 
SFR estimates are presented in Table~\ref{t:sfr}.

\begin{deluxetable}{lcc}
\tabletypesize{\scriptsize}
%  \tabletypesize{\normalsize} 
\tablecaption{Outflowing MgII Absorption in Star Forming Cluster Members} \tablewidth{0pt} \tablehead{ 
    \colhead{Galaxy} 
    & \colhead{MgII}
    & \colhead{Velocity Relative}
%    & \colhead{SFR} 
     \\
     \colhead{ID} 
    & \colhead{Redshift} 
    & \colhead{to [O II] (km s$^{-1}$)}%
%    &  \colhead{M$_{\odot}$ yr$^{-1}$} 
    } \\
    \startdata
J204110.1-444933.6  & 1.4718 &  $-$510 $\pm$ 160 \\
J204100.1-445025.2  & 1.4758 &  $-$230 $\pm$ 80 \\
J204100.9-445315.7  & 1.4764 &  $-$250 $\pm$ 70 \\
J204058.1-445206.7  & 1.4768 &  $-$20 $\pm$  60 \\
J204051.2-445116.8  & 1.5097 &  $-$280 $\pm$ 60 \\
J204050.3-445020.5  & 1.4720 &  $-$970 $\pm$ 80  \\
\hline
\enddata
\label{t:mgii}
%\tablenotetext{a}{[O II] flux measured within $\pm$ 2-$\sigma$ of the line centroid, uncorrected for 
%slit losses, which should be very small for objects that were the primarily targets of individual mask 
%slits.}
\end{deluxetable}

\subsection{SZ Mass Estimate}

We update the SZ mass estimate from \citet{reichardt13}, incorporating the 
newly measured spectroscopic redshift for \cluster. The SZ mass is calculated 
using a Markov chain Monte Carlo (MCMC) method that fits the SZ 
mass-observable scaling relations while marginalizing over $\Lambda$CDM 
cosmological parameters, and incorporates constraints 
available from X-ray data for 14 SPT clusters, as well as observations 
of the Cosmic Microwave Background, the cosmic baryon density measured from 
primordial deuterium abundance, baryon acoustic oscillations, distance 
measurements from type Ia supernovae, and the galaxy cluster mass function as 
measured by the SPT. The MCMC method is described in more detail in 
\citet{reichardt13} and \citet{benson13}. The resulting mass is defined as the 
mass within a radius, r$_{500}$, within which the cluster has a mean matter 
density that is 500 times the critical density of the universe, $\rho_{c}$(z), and 
is calculated to be M$_{500,SZ} =$ 3.2 $\pm$ 0.8 $\times$ 10$^{14}$ M$_{\odot}$ 
h$_{70}^{-1}$. This mass estimate includes measurement noise, noise due to 
astrophysical contaminants, and the systematic errors due to the uncertainties 
in scaling relation parameters and cosmological parameters. It is also common 
to report galaxy cluster masses within the radius 
r$_{200}$ which encloses a region that is 200 times the \emph{mean} density of the 
universe; assuming the NFW profile shape \citep{navarro97} for the cluster 
density profile and using a value for the concentration parameter taken from the 
mass-concentration relation as measured in simulations \citep{duffy08}, it 
becomes straightforward to convert between masses measured at different 
over-density radii \citep{hu03a}. The r$_{200}$ SZ-based mass estimate for 
\cluster~is M$_{200,SZ} =$ 5.8 $\pm$ 1.4 $\times$ 10$^{14}$ 
M$_{\odot}$ h$_{70}^{-1}$.

The existence of massive galaxy clusters at relatively early epochs of the universe 
has the potential to test the viability of cosmological models, and with both a mass 
and redshift in-hand for \cluster~we can quantify its rarity (or lack thereof). 
Following the procedure in Section 4.1 of Stalder et al. (2013), we can estimate how 
many clusters at least as rare as \cluster~ that we would expect in the SPT-SZ survey. 
Given the best-fit mass function and scaling relation from \citet{reichardt13}, we expect 
approximately 0.7 clusters with simultaneously higher mass and redshift than \cluster.  
If we consider an ensemble of 720 deg$^{2}$ of SPT-SZ survey area 
\citep[i.e., the sample in][]{song12} then we find that we are very likely ($>$99\%) to 
have found a cluster at least as rare as \cluster. Running the same test for the full 
2500 deg$^{2}$ SPT-SZ survey area we naturally also find that it is very likely that we 
should ($>$99\%) find a cluster at least as rare as \cluster. 

\begin{deluxetable*}{lcrcc}
\tabletypesize{\scriptsize}
%  \tabletypesize{\normalsize} 
\tablecaption{Emission Line properties of Confirmed Cluster Members} \tablewidth{0pt} \tablehead{ 
    \colhead{} 
    & \colhead{R$_{proj}$} 
    & \colhead{v$_{peculiar}$}
    & \colhead{[O II]\tablenotemark{a} (\scriptsize{$\times$10$^{-17}$})}
    & \colhead{SFR$_{[O~II]}$} 
     \\
     \colhead{ID} 
    & \colhead{Mpc h$_{70}^{-1}$} 
    & \colhead{km s$^{-1}$}
    & \colhead{\scriptsize{ erg cm$^{-2}$ s$^{-1}$}}
    &  \colhead{M$_{\odot}$ yr$^{-1}$} 
    } \\
\startdata
J204110.1-444933.6  &  1.454 & $-$240 $\pm$ 40 &  3.99 $\pm$ 1.04 & 8.0 $\pm$ 3.2  \\
J204100.1-445025.2  &  0.607 & $-$30 $\pm$ 40 &  1.27 $\pm$ 0.61 & 2.6 $\pm$ 1.4 \\
J204057.0-445213.7\tablenotemark{b}  &  0.384  & 750 $\pm$ 40 &  $>$1.63 $\pm$ 0.51  & $>$3.3 $\pm$ 1.4  \\
J204100.9-445315.7  &  0.878 & $-$180 $\pm$ 40  &  4.96 $\pm$ 1.22 & 10.0 $\pm$ 3.9 \\
J204057.2-445121.4\tablenotemark{b}  &  0.222  & $-$2900 $\pm$ 40 &  $>$3.66 $\pm$ 0.90 & $>$7.1 $\pm$ 2.7  \\
J204057.3-445108.6\tablenotemark{b}  &  0.292  & $-$1050 $\pm$ 60 &  $>$0.81 $\pm$ 0.37 & $>$1.6 $\pm$ 0.9  \\
J204113.6-445125.2  &  1.306  & $-$3270 $\pm$ 30 &  5.97 $\pm$ 1.36 & 11.5 $\pm$ 4.3  \\
J204058.1-445206.7  &  0.283  & 110 $\pm$ 30 &  11.5 $\pm$ 2.36 & 23.1 $\pm$ 8.4 \\
J204054.6-445201.1\tablenotemark{b}  &  0.472 & 760 $\pm$ 40   &  $>$1.90 $\pm$ 0.58 & $>$3.9 $\pm$ 1.7 \\
J204051.2-445116.8  &  0.750  & 4100 $\pm$ 40 &  1.73 $\pm$ 0.59  & 3.7 $\pm$ 1.7  \\
J204050.3-445020.5\tablenotemark{b}  &  1.036 & 250 $\pm$ 40  &  $>$0.82 $\pm$ 0.41 & $>$1.7 $\pm$ 1.0 \\
J204048.5-445021.4\tablenotemark{c}  &  1.162 & 640 $\pm$ 40  &  $>$1.47 $\pm$ 0.53 & $>$2.9 $\pm$ 1.4 \\
J204044.2-445124.0  &  1.364 & 30 $\pm$ 40  &  0.72 $\pm$ 0.47 & 1.5 $\pm$ 1.0 \\
J204050.4-445022.2  &  1.015  & $-$260 $\pm$ 40 &  2.22 $\pm$ 0.69 & 4.5 $\pm$ 1.9 \\
J204048.7-445020.7\tablenotemark{c}  &  1.148 & 340 $\pm$ 40 &  $>$7.90 $\pm$ 1.68 & $>$16.0 $\pm$ 5.9 \\
\hline
\enddata
\label{t:sfr}
\tablenotetext{a}{[O II] flux measured within $\pm$ 2-$\sigma$ of the line centroid, uncorrected for 
slit losses, which should be very small for objects that were the primarily targets of individual mask 
slits.}
\tablenotetext{b}{These galaxies fell serendipitously onto slits, and therefore likely suffered 
significant slit losses that are difficult to quantify robustly, so we report the measured [O II] flux as 
a lower limit.}
\tablenotetext{c}{These objects appear as a blend of two sources in the IRAC catalogs that were 
used to design our spectroscopic masks, such that the mask slit falls partially onto both sources. 
As a result we avoid attempting an ad hoc correction for slit losses and report the measured [O II] 
fluxes and SFRs as lower limits.}
\end{deluxetable*}

\subsection{Velocity Dispersion}\label{ss:vdisp}

Our ability to compute a reliable velocity dispersion is fundamentally 
hindered by the small number of available cluster member velocities. However, 
given the paucity of spectroscopically confirmed members in known high redshift 
galaxy clusters, the spectroscopy presented here for \cluster\ represents one of the 
best-sampled velocity distributions for a galaxy cluster at $z > 1.2$. It is therefore 
interesting to investigate the dynamics of \cluster, while keeping in mind the caveat 
that the sample used is limited to 15 spectroscopic galaxies.

\begin{figure}
\begin{center}
\epsscale{1.21}
%\epsscale{0.9}
\plotone{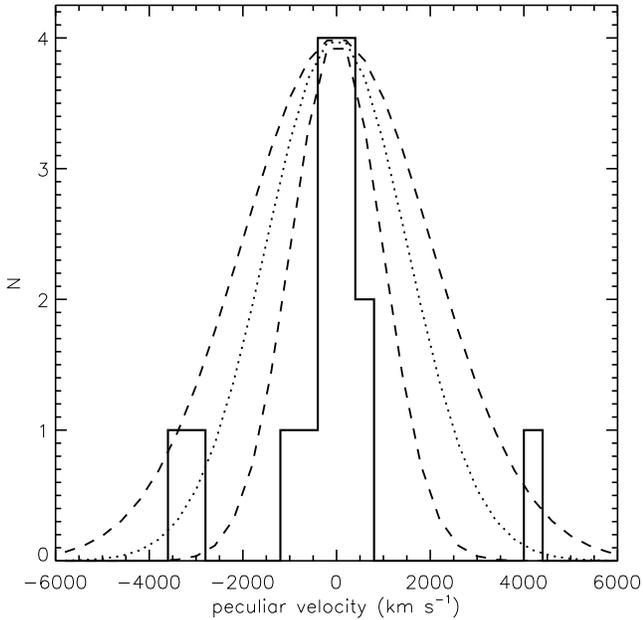}
\caption{ The rest-frame 
velocity histogram for \cluster, with peculiar velocities grouped into bins of 400 
km s$^{-1}$ and the estimate of the velocity dispersion over-plotted 
(dotted line), along with the velocity distributions corresponding to the jackknife 
1-$\sigma$ uncertainties (dashed lines) in the dispersion.}
\label{vhist}
\end{center}
\end{figure}

The bi-weight estimate of the median redshift for \cluster~is $z = 1.478^{+0.003}_{-0.003}$, 
and we compute the velocity dispersion from the sample of 15 cluster members 
with emission line redshifts using a gapper statistic similar to that described by 
\citet{beers90}. The bi-weight estimator is commonly used in the literature to 
measure the dispersion in peculiar velocities of cluster members, but \citet{beers90} 
point out 
that the gapper statistic is more robust for sparsely sampled distributions (e.g., 
N $\leq$ 15) so we use the gapper estimate in this work to produce the most 
reliable estimate. The velocity dispersion of \cluster~is $\sigma_{v,gap}$ 
$=$ 1500 $\pm$ 520 km s$^{-1}$, where the uncertainties are computed using the 
jackknife method. For reference, both the bi-weight and simple standard deviation 
estimates of the dispersion for \cluster~(1600 and 1660 km s$^{-1}$, respectively) 
are in reasonable agreement with the gapper value. The velocity distribution is 
shown in Figure~\ref{vhist}, along with the estimated distribution and its 
jackknife uncertainties. 

We note that of the 15 spectroscopic cluster members, there are two pairs of cluster 
members that are separated by small ($\lesssim$3\arcsec; $\lesssim$ 30 kpc h$_{70}^{-1}$) 
angular distances on the sky (second and third cutouts from the top on the right side of 
Figure~\ref{f:2dspec}). Each of these pairs corresponds a slit on our custom 
spectroscopic slit-masks that yield spectral traces for two different galaxies at the 
cluster redshift. From the data we know that these galaxies are located close together 
in both projected distance on the sky, and in recession velocity. There are several possible 
physical interpretations of these pairs: 1) they could be two cluster member galaxies that 
appear as a chance projection (i.e., galaxies within the cluster that are separated by a large 
distance in the radial direction), 2) they could be cluster member galaxies that are physically 
close to one another, or 3) each pair could in fact be [O II] emission from two star forming 
regions within a single galaxy. In each case the pairs of galaxies are separated in velocity 
space by dv $>$ 500 km s$^{-1}$, which allows us to rule out the possibility that each pair is 
really just two different star forming regions within the same galaxy. Given the limited phase 
space information we cannot measure the true phase space coordinates of each galaxy pair, 
however, and therefore we cannot distinguish between the first and second possibilities 
above. If one or both of these galaxy pairs are, in fact, located physically close to one 
another then it is possible that they are parts of some subhalo/substructure within the 
larger cluster potential. If this is the case then these galaxies are not 
all necessarily providing independent samplings of the total cluster potential. 
There is possible evidence for this sub-halo sampling in the velocity distribution, 
shown in Figure~\ref{vhist}, where there is a concentration of cluster member 
velocities with a small dispersion and three more outlying galaxies. We also note 
that three of the 15 galaxies that we have identified as members have very large 
peculiar velocities ($\gtrsim$3000 km s$^{-1}$) relative to the bi-weight median, 
and one or more of these could be interlopers in the velocity distribution, but are 
not rejected by 3-$\sigma$ cuts with the current 15 member velocity sample.

The dispersion estimated from the 15 members is poorly constrained, with a 
2-$\sigma$ range of $\sim$500-2500 km s$^{-1}$. If we apply the scaling relation 
between velocity dispersion and virial mass of \citet{evrard08} then this corresponds 
to a mass of M$_{200,\sigma}$ $=$ 1.8$^{+2.5}_{-1.3}$ $\times$ 10$^{15}$ 
M$_{\odot}$ h$_{70}^{-1}$, which is both extremely large and extremely uncertain. 
However, it is also physically unreasonable to expect the velocity dispersion 
estimated here to be related to the cluster mass in the same way as that of a 
dynamically relaxed population of galaxies (e.g., passive, early type). All 15 
spectroscopically confirmed members of \cluster\ exhibit strong [O II] emission, and 
numerous studies have empirically confirmed that the velocity dispersions measured from 
blue/late-type/star forming galaxies in clusters -- which tend to be infalling -- are larger than 
the dispersion of their passive counterparts 
\citep{girardi96,mohr96,carlberg97b,koranyi00,goto05,pimbblet06}. 
Studies using simulations similarly find that cluster velocity dispersions measured using blue 
galaxies are larger than those measured from red galaxies \citep{gifford13a}.

It is, therefore, interesting to proceed with the hypothesis that our sample of spectroscopic 
cluster members are all in-falling, and may be treated as test particles falling into the cluster 
potential. In this scenario, their line of sight velocities distribution would reflect the free fall 
velocity, rather than the velocity dispersion, associated with the cluster mass. As mentioned 
above, it has been shown in observations and simulations that cluster velocity dispersions 
computed using blue galaxies are, on average, systematically larger than 
those computed from red/passive galaxies, which qualitatively affirms a physical scenario 
in which the distribution of blue galaxy velocities trace the cluster potential via infalling rather 
than virialization.  The equation 
relating the free fall velocity, v$_{\rm ff}$, to the attracting mass, assuming the galaxies began 
falling from some distance, $r >> R$, is $$M(<R) \simeq \frac{~Rv_{\rm ff}^{2}}{2G},$$

\noindent where R is the current distance between an infalling galaxy and the center of mass 
of the cluster. We can use the projected distance of the galaxies from the cluster 
center (Table~\ref{t:sfr}) -- assuming that their trajectories are randomly oriented on 
the sky -- to estimate the median distance from the 15 member 
galaxies to the center of \cluster. The median projected distance on the sky, $\bar{R}_{proj}$, 
of the 15 member galaxies is 0.88 Mpc (corresponding to $\bar{R} =1.8$ Mpc after de-projection 
assuming velocity vectors randomly oriented on the sky), and solving for the mass here gives 
M($<$ $\bar{R}$) $\simeq$ 5 $\times$ 10$^{14}$ M$_{\odot}$ h$_{70}^{-1}$, consistent 
with the mass estimated from the SZ signal. We do not advocate for this method as a 
way to precisely estimate cluster masses, but we do note that the results of  this free-fall 
picture are consistent with the SZ mass estimate, and makes sense in the context of a 
physical picture in which the 15 [O II] emitting galaxies are predominantly in-falling cluster 
member galaxies.

\section{Discussion}\label{s:discussion}

\subsection{Cluster Members In Color-Magnitude Space}

\begin{figure}
\begin{center}
\epsscale{1.21}
%\epsscale{0.58}
\plotone{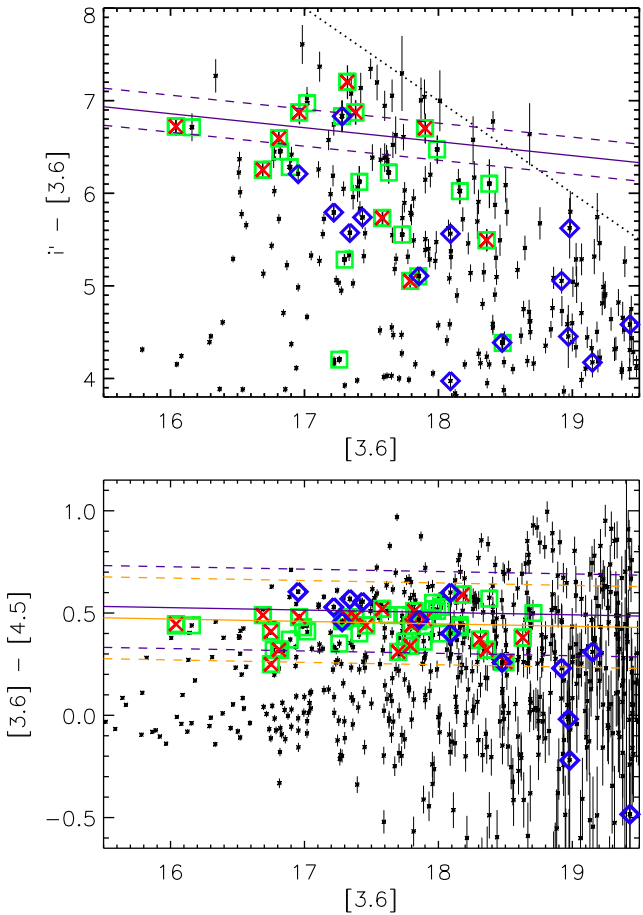}
\caption{
Color-magnitude diagrams for \cluster~using the best available optical 
and NIR photometry -- Megacam $i$' band, plus [3.6] and [4.5] from 
Spitzer$+$IRAC. All sources that have detections in the IRAC and Megacam imaging 
(i.e., essentially everything within an area on the sky set by the IRAC field of view) 
are plotted as small black asterisks, with candidate cluster members identified from 
the Spitzer photometry over-plotted as green squares. Spitzer-selected cluster 
candidates that were targeted by spectroscopic slits in our Sept 2012 IMACS MOS 
observations but did not yield a redshift are also identified by red X's, and cluster 
members confirmed via [O II] emission are marked with blue diamonds. Galaxies 
plotted as empty green 
squares were photometrically identified cluster members that did not receive 
a slit in the mask design process. $Top:$ 
$i$'$-$[3.6] vs [3.6]. The diagonal dotted line indicates the color-magnitude 
selection corresponding to the 50\% completeness limit in our $i$'-band photometry. 
Over-plotted in purple is the red sequence from \citet{bruzual03} models with a 
Chabrier initial mass function \citep{chabrier03} and solar metallicity, assuming a 
single burst of star formation at $z = 5$.
$Bottom:$ [3.6]$-$[4.5] vs [3.6] with 
\citet{bruzual03} models -- $\pm$0.2 mags in color -- at $z = 1.41$ (photometric 
cluster redshift; orange) and $z = 1.478$ (purple).}
\label{cmds}
\end{center}
\end{figure}

In Figure~\ref{cmds} we plot the results of our spectroscopy on top of the 
optical$+$NIR $i$'$-$[3.6] vs [3.6] and IRAC NIR only [3.6]$-$[4.5] vs [3.6] 
color-magnitude diagrams (CMDs) for \cluster.  The IRAC-only CMD is 
useful for identifying an over-density of galaxies in redshift based on the 
presence of the rest-frame 1.6$\mu m$ ``stellar bump'' feature that is 
ubiquitous in older stellar populations with similar ages and formation 
histories \citep[e.g.,][]{brodwin06,muzzin13}, while the optical$+$IRAC 
CMD is sensitive to red/passively evolving galaxies at a common redshift. 
Object prioritization for spectroscopic mask design was based primarily on 
the \emph{Spitzer} CMD, and it is clear that the objects which form a tight 
sequence in [3.6]$-$[4.5] vs [3.6] are not tightly clustered in 
$i$'$-$[3.6] vs [3.6] space -- i.e., they are not a monolithic passively 
evolving population of galaxies. The spectroscopically identified star forming 
cluster members tend to occupy the ``blue cloud'' region in the $i$'$-$[3.6] vs 
[3.6] CMD, as expected.

In addition to the population of actively star forming galaxies revealed in our 
spectroscopy, there is also possible evidence for a significant population of passive cluster 
members. Their presence can be inferred from the extremely low S/N continuum 
emission that we observe in MOS slits placed on photometrically selected cluster 
member galaxies. There are 20 such objects plotted as red X's in Figure~\ref{cmds}, 
though only 10/20 have $i$'-band detections. Those without cannot be included in the 
$i$'$-$[3.6] vs [3.6] CMD. We cannot claim that these 20 galaxies are all passive cluster 
members, but it is unlikely that most or all of them are interloping passive 
galaxies given that they have IRAC colors that are consistent with a population of 
galaxies at the spectroscopic redshift of \cluster. It is also encouraging 
that half of these putative passive member galaxies with $i$'-band detections fall 
within 0.2 magnitudes of the red sequence predicted for a population of passively 
evolving galaxies at the cluster redshift. Deeper spectroscopic 
observations, preferably using the nod-and-shuffle technique in the optical, or one 
of the new generation of multi-object NIR spectrographs, will be necessary to 
unambiguously identify passive member galaxies of \cluster.

The ground-based NIR imaging that is currently available for \cluster~is not 
sufficiently deep to allow us to construct a CMD which narrowly brackets the 
4000\AA~break in order to isolate passive cluster members e.g., $i$'$-J$ vs $J$ or 
$i$'$-K_{s}$ vs $K_{s}$. Deeper NIR imaging in the $\sim$1-2.3$\micron$ range 
would allow us to identify the red sequence, and facilitate a measurement of the 
luminosity function of the passive galaxy population. 

\subsection{Prevalence of Star Forming Cluster Members}

The abundance of strong [O II] emitting galaxies 
in \cluster~stands in stark contrast to other spectroscopically confirmed $z > 1$ SPT 
clusters, and is consistent with the model discussed above, likely reflecting both its 
lower mass and higher redshift relative to the majority of the SPT cluster sample, 
which are mass selected to satisfy M$_{500,SZ} > 3 \times 10^{14}$~h$_{70}^{-1}$ 
M$_{\odot}$, and have a median redshift of z $=$ 0.55. Given the incompleteness of 
our spectroscopic coverage (i.e., we do not have spectroscopy of a magnitude limited 
sample), it is difficult to quantify the abundance of star forming members in an absolute 
sense. However, what we {\it can} do is compare the abundance of star forming members 
in \cluster~relative to other high-redshift cluster that were observed with the same 
spectroscopic strategy (slit placement and object prioritization) as the data presented 
in this paper.

The IMACS observations presented here included 
a total of 59 slits, resulting in 15 [O II] emitters (i.e., a ``hit rate'' of 25.4 $\pm$ 7\%). 
These IMACS observations result from masks designed using the same data as 
an input  -- primarily {\it Spitzer}/IRAC photometry -- and using the same object 
selection criteria as SPT-CL~J0546-5345, SPT-CL~J2106-5844, and SPT-CL~J0205-5829, 
at $z = $1.067, 1.132, and 1.320, respectively \citep{brodwin10,foley11,stalder13}. 
The IRAC photometry is sensitive to galaxies deep down the luminosity function at all 
of these redshifts (M$^{*} + 2.5$ at z $\sim$ 1.5), such that IRAC color-based selections 
are not biased, e.g., picking out only the brightest cluster members at higher redshifts. 
Additionally, all of these high-z SPT clusters were observed with equivalent 
wavelength coverage and spectral resolution,  similar integration times, and in 
similar conditions. Observations of each of these three previously published systems 
resulted in $\leq$ 3 emission line cluster member galaxies per cluster, and we can 
compute the hit rate for  [O II] emitting cluster members resulting from spectroscopic 
slits. The resulting hit rates are 4 $\pm$ 3\%, 2 $\pm$ 2\%, and 2 $\pm$ 2\% for 
SPT-CL~J0546-5345, SPT-CL~J2106-5844, and SPT-CL~J0205-5829, respectively. 
Additionally, the emission line cluster members in these three previously published 
SPT-discovered clusters exhibit less observed [O II] flux than the star forming 
galaxies in \cluster. 

The identification here of 15 strong emission line galaxies implies that 
\cluster~is experiencing a period of star formation that far exceeds the other 
spectroscopically studied SPT galaxy clusters at $z > 1$. The discovery of 
\cluster~at $z = 1.478$ marks the first SZ-detected galaxy cluster 
to be observed in an epoch in which even moderately massive clusters (e.g., 
$\sim$5 $\times$ 
10$^{14}$ M$_{\odot}$) have not yet settled into the mode of passive evolution 
that is associated with massive, evolved clusters at lower redshift.

As indicated in Figure~\ref{cmds} and discussed in Section~\ref{ss:members}, there 
is also some evidence for a population of passive members in \cluster, for which our 
instrument and spectroscopic setup were not well-suited to measure redshifts. 
Further multi-wavelength observations of \cluster~will be necessary to fully 
characterize the passive and star forming galaxy populations, but the significant 
abundance of strong [O II] emitting galaxies revealed by our spectroscopy is a strong 
indication that this galaxy cluster is undergoing significant build-up of new stellar mass, 
similar to other high redshift clusters discovered at other wavelengths.

Detailed studies of other high redshift clusters find evidence for assembly of cluster 
member galaxies through increased merging activity. For example, 
in \emph{HST} imaging of ClG J0218.3-0510 at z = 1.62, \citet{lotz12} observe a high 
incidence of double-nuclei galaxies and close galaxy pairs in candidate cluster 
member galaxies with large stellar mass ($\gtrsim$3 $\times$ 10$^{10}$ M$_{\odot}$), 
from which they infer a merger rate as much as an order of magnitude higher than in 
similarly massive field galaxies at the same redshift. \citet{tran10} also measure a high 
star formation density ($\sim$ 1700 M$_{sun}$ yr$^{-1}$ Mpc$^{-1}$) in 
ClG J0218.3-0510 at z = 1.62 using a combination of 10-band SED fitting and 
spectroscopy.  \citet{rudnick12} use 
the measured luminosity function (LF) of red sequence members in ClG J0218.3-0510 
to argue that increased mergers are necessary to describe the build up of galaxies 
in clusters. \citet{zeimann12} also find high star formation rates traced by rest-frame 
nebular emission lines in spectroscopically confirmed members of IDCS J1433.2+3306 
-- an optical$+$NIR selected cluster at $z = 1.89$. 

Looking beyond individual high-z clusters, several groups have also measured the 
properties of cluster member galaxies in larger samples of high-redshift galaxy clusters. 
{\it Spitzer}/IRAC imaging can be used to identify color-selected cluster member galaxies 
based on the 1.6$\mu m$ bump feature, which can identify galaxies that are passive, or 
actively star forming, or in the processes of transition from one the other. 
\citet{mancone10} measure {\it Spitzer}/IRAC [3.6] and [4.5] LFs for binned samples of 
optical$+$NIR selected galaxy clusters, and find disagreement between the measured 
LF in $z \gtrsim 1.3$ clusters and the assumed passive evolution model, which they 
suggest could be evidence for ongoing galaxy mass assembly. Similarly, in a sample 
of 16 \emph{Spitzer}-selected clusters, \citet{brodwin13} combine {\it Spitzer} MIPS, 
IRAC, optical, and spectroscopic data to characterize the formation histories of cluster 
member galaxies; they show evidence for a systematic increase in star 
formation at $z \gtrsim 1.4$, and propose a model in which galaxy clusters undergo 
an epoch of frequent merging activity that resembles group environments in the local 
universe \citep{hopkins08}. In this model, merger activity falls off steeply as clusters 
become more relaxed, with larger internal velocity dispersions.

%%%%%%%%%%%%%%%%%%%
%%  Conclusions  %%
%%%%%%%%%%%%%%%%%%%

\section{Summary and Conclusions}\label{s:conc}

We present the discovery and follow-up observations of \cluster, with a 
spectroscopic redshift of $z = 1.478^{+0.003}_{-0.003}$. It is the highest-redshift, 
spectroscopically-confirmed, SZ-discovered cluster known. We combine the 
newly measured redshift with SPT observations to infer a mass of M$_{500,SZ}$ 
(M$_{200}$) $=$ 3.2 $\pm$ 0.8 (5.8 $\pm$ 1.4) $\times$ 10$^{14}$ M$_{\odot}$ 
h$_{70}^{-1}$, making \cluster\ one of the most massive clusters known at $z > 1.4$.
We estimate the cosmological rarity of SPT-CL J2040-4451, and find that it is not 
surprising to find a cluster of this mass and redshift in the SPT-SZ survey.  

From our optical spectroscopy we identify 15 cluster members with 
[O II]$\lambda$$\lambda$ 3727 emission, all of which exhibit star formation 
rates $\geq 1.5$ M$_{\odot}$ yr$^{-1}$. The abundance of star forming galaxies 
observed in \cluster~relative to other high-z SPT-detected clusters agrees well with 
recent observations that reveal elevated star formation in galaxy clusters at $z \gtrsim 1.4$. 
We measure a velocity dispersion 
of the star forming cluster members of $\sigma_v = 1500 \pm 520$ km s$^-1$.  However, 
we argue that this measurement is likely biased high, relative to the expectation from 
the dark matter halo mass, due to the fact that all of the measured cluster members are 
star-forming, and therefore more likely to be drawn from the population of galaxies that 
are in-falling into the cluster, rather than the dynamically relaxed population of passive 
cluster member galaxies.

Notably, \cluster~is not only the highest redshift cluster 
in the current SPT catalog, but it is also near the low end of the mass range that 
the SPT-SZ survey samples. Studying the epoch of star formation in the 
progenitors of the most massive galaxy clusters requires that we investigate cluster 
assembly as a function of both redshift {\it and} mass. It is therefore important to 
use samples of high redshift clusters that can be precisely classified as a function of 
mass. A significant advance towards this goal has been achieved with the recently 
completed 2500 deg$^2$ SPT-SZ survey, which provides a nearly-mass-independent 
cluster catalog out to arbitrarily high-redshift. This catalog 
contains $ N \gtrsim 30$ clusters at $z > 1$, providing the largest, mass-selected 
cluster sample at these redshifts. A dedicated study 
of this sample will help to place \cluster~in context with respect to the star 
forming activity in the most massive high redshift clusters.

\begin{acknowledgments}

 {\it Facilities:} \facility{Blanco (MOSAIC2)},
  \facility{Blanco (NEWFIRM)},
 \facility{Magellan:Baade (IMACS)},
  \facility{Magellan:Clay (MegaCam)},
  \facility{Magellan:Baade (FourStar)},
 \facility{Spitzer (IRAC)},
 \facility{South Pole Telescope}

~~~~~~

The South Pole Telescope program is supported by the National Science
Foundation through grant ANT-0638937.  Partial support is also
provided by the NSF Physics Frontier Center grant PHY-0114422 to the
Kavli Institute of Cosmological Physics at the University of Chicago,
the Kavli Foundation, and the Gordon and Betty Moore Foundation.
Galaxy cluster research at Harvard is supported by NSF grant AST-1009012.  
Galaxy cluster
research at SAO is supported in part by NSF grants AST-1009649 and
MRI-0723073. The McGill group acknowledges funding from the National
Sciences and Engineering Research Council of Canada, Canada Research
Chairs program, and the Canadian Institute for Advanced Research.
X-ray research at the CfA is supported through NASA Contract NAS
8-03060. The Munich group acknowledges support from the Excellence
Cluster Universe and the DFG research program TR33.
This work is based in part on observations obtained with the Spitzer Space
Telescope (PID 60099), which is operated by the Jet Propulsion
Laboratory, California Institute of Technology under a contract with
NASA. Support for this work was provided by NASA through an award
issued by JPL/Caltech.  Additional data were obtained with the 6.5~m
Magellan Telescopes located at the Las Campanas Observatory,
Chile and the Blanco 4~m Telescope at Cerro Tololo Interamerican
Observatories in Chile.
R.J.F.\ is supported by a Clay Fellowship.  B.A.B\ is supported by a KICP
Fellowship, M.Bautz acknowledges support from contract
2834-MIT-SAO-4018 from the Pennsylvania State University to the
Massachusetts Institute of Technology. M.D.\ acknowledges support
from an Alfred P.\ Sloan Research Fellowship, W.F.\ and C.J.\
acknowledge support from the Smithsonian Institution. This research 
used resources of the National Energy Research Scientific Computing 
Center, which is supported by the Office of Science of the U.S. Department 
of Energy under Contract No. DE-AC02-05CH11231.

\end{acknowledgments}

\bibliographystyle{fapj}
\bibliography{spt2040.bib}

%\eject

\end{document}